\documentclass[12pt]{article}

\usepackage[totalheight = 23cm, totalwidth = 17cm]{geometry}
\usepackage{amssymb,amsmath,amsfonts,amsbsy,graphicx,bm}

\usepackage{color,cancel,ulem, hyperref}

\newcommand{\dis}[1]{\begin{equation}\begin{split}#1\end{split}\end{equation}}

\begin{document}

\begin{titlepage}

\begin{center}

{\setlength{\baselineskip}%
{1.5\baselineskip}
{\LARGE \bf 
Bounds on complex structure moduli values for perturbative control 
}
\par}

\vskip 1.0cm

{\large
Min-Seok Seo$^{a}$ 
\footnote[0]{e-mail : minseokseo57@gmail.com}
}

\vskip 0.5cm

{\it
$^{a}$Department of Physics Education, Korea National University of Education,
\\ 
Cheongju 28173, Republic of Korea
}

\vskip 1.2cm

\end{center}

\begin{abstract}

 String compactification in the framework of the low energy effective supergravity requires the perturbative control in both the large volume and the weak coupling expansions.
 However, when the complex structure moduli couple to some lattice structure, the Sp$(2(h^{2,1}+1))$ symmetry of the tree level K\"ahler potential allows the correction to the K\"ahler potential to diverge in the large field limit of the complex structure moduli, resulting in the breakdown of the perturbative control.
 Here the lattice structure naturally appears in the presence of a tower of states like the Kaluza-Klein or the string modes, an essential ingredient of the distance conjecture.
 The similar situation can be found from the axio-dilaton contribution to the corrected K\"ahler potential, where the SL$(2, \mathbb{Z})$ symmetry as well as the coupling between the axio-dilaton and the lattice structure allow  the correction to diverge in the weak coupling limit.
 In order to keep the perturbative control, the values of the complex structure moduli as well as the dilaton  must have the upper bound, which is determined by the volume of the internal manifold and the string coupling constant, hence the Kaluza-Klein and the string mass scales. 
 The form of the bounds are quite similar to that given by   the distance conjecture, both prevents the descent of a tower of states.

\end{abstract}

\end{titlepage}

\newpage

\section{Introduction}

Whereas the 10-dimensional critical string theory is a well-motivated candidate for the UV completion of quantum gravity, we need an additional process, compactification, to make contact with the 4-dimensional spacetime of everyday experience (for reviews, see, e.g., \cite{Ibanez:2012zz, Tomasiello:2022dwe}).
This allows us to explain various features of the resulting 4-dimensional theory in terms of the size and shape of the 6-dimensional internal manifold, which are determined by the stabilized values of the K\"ahler and the complex structure moduli, respectively.
Typically, the dynamics of these moduli are studied in the framework of low energy effective supergravity.
In order to derive the effective supergravity action from string theory, two limits must be taken :
\begin{itemize}
\item  To obtain appropriate equations of motion from the vanishing condition of the world-sheet $\beta$-functions,  the sigma model loop expansion parameter $\alpha'=\ell_s^2/(2\pi)^2$ is required to be much smaller than the typical length scale of the background geometry.
From this  the string excitations are expected to decouple from the low energy effective field theory (EFT).
In particular, the value of ${\cal V}$, the string frame volume of the internal manifold in units of the string length, must be large.
 \footnote{We note that the {\it string frame} volume ${\cal V}$  is  related to the {\it Einstein frame} volume ${\cal V}_E$ by ${\cal V}=g_s^{3/2}{\cal V}_E$ (since $G_{MN}^{(E)}=e^{-\Phi/2}G_{MN}^{(s)}$). 
 Whereas ${\cal V}_E$ is often used to obtain the physical quantities in the low energy physics, the expansion parameter consistent with the $\alpha'$-expansion is ${\cal V}$, rather than ${\cal V}_E$ (see, e.g., \cite{Conlon:2005ki}).}
This  large volume limit ${\cal V}\to \infty$ is equivalent to the asymptotic limit of the K\"ahler moduli where the field value of the scalar part is extremely large.  
\item The effective supergravity action also can be constructed from the string scattering amplitudes in the limit $\alpha'\to 0$.
In the string loop expansion of them, since the tree-level contributions are equivalent to those obtained from the supergravity theory, it is additionally required that the string loop corrections are suppressed.
This   is achieved by taking  the weak coupling limit $g_s \ll 1$.
Here the string coupling constant $g_s$ is given by $e^{\langle \Phi\rangle}$, the exponential of the dilaton vacuum expectation value (VEV).
In type IIB supergravity we are interested in, one can define the axio-dilaton field $\tau =C_0 +i e^{-\Phi}$ ($C_0$ : the Ramond-Ramond (RR) 0-form), then the weak coupling limit becomes the asymptotic limit Im$\tau \to \infty$.
\end{itemize}

 As such,   both limits for the supergravity approximation correspond to the asymptotic regimes in the field space of the K\"ahler moduli and the axio-dilaton.
However, the field values cannot be arbitrarily large.
 On one hand, as ${\cal V} \to \infty$,  the Kaluza-Klein (KK) modes become light, so the low-energy EFT is no longer 4-dimensional.
On the other hand, while the geometry of the internal manifold decouples from the string excitations in this limit, 4-dimensional gravity does not. 
That is, since the squared Planck length is given by
\dis{\kappa_4^2=\frac{1}{M_{\rm Pl}^2}=\frac{\kappa_{10}^2}{{\rm Vol}(X_6)}=\frac{g_s^2}{4\pi}\frac{\ell_s^2}{{\cal V}},}
where $\kappa_{10}^2=\frac{g_s^2}{4\pi}(4\pi^2\alpha')^4 =\frac{g_s^2 }{4\pi}\ell_s^8$,
  or equivalently, 
 the string mass scale satisfies
\dis{M_s^2 =\frac{1}{\ell_s^2}=\frac{g_s^2}{4\pi{\cal V}}M_{\rm Pl}^2\label{eq:Ms},}
 the ratio $\ell_s/\kappa_4$ is larger than $1$ in the supergravity approximation, and in particular, diverges when either ${\cal V} $ or Im$\tau = 1/g_s$   is extremely large.
In other words, the string mass scale $M_s$ giving a fixed value of $M_{\rm Pl}=2.43\times 10^{18}$ GeV gets lower toward the asymptotic regimes.
Regarding this, the distance conjecture \cite{Ooguri:2006in} claims that the EFT is invalidated in the asymptotic regime of the field space due to the descent of a tower of states, the KK or the string tower \cite{Lee:2019xtm, Lee:2019wij}.
 The conjecture can be extended to the complex structure moduli space : from studies based on the asymptotic Hodge theory \cite{Grimm:2018ohb, Grimm:2018cpv} (see \cite{Schmid:1973cdo, Carrani:1986} and also \cite{Grimm:2019ixq}), it turns out that a tower of BPS states generated by D-branes wrapping the cycles becomes extremely light in the asymptotic regime.
 
Meanwhile, it was conjectured that the flux cannot stabilize all the complex structure moduli  \cite{Bena:2020xrh, Bena:2021wyr}, contrary to na\"ive expectation \cite{Dasgupta:1999ss, Giddings:2001yu} (for reviews on the flux compactification, see, e.g., \cite{Grana:2005jc, Douglas:2006es, Blumenhagen:2006ci}).
 The numerical tests as well as the analysis grounded in the asymptotic Hodge theory show that this tadpole conjecture seems to hold in the strict asymptotic regime of the complex structure moduli space (see, e.g., \cite{Plauschinn:2021hkp, Lust:2021xds, Grimm:2021ckh, Grana:2022dfw, Tsagkaris:2022apo}).
However, the situation changes in the interior of the moduli space, where the flux may  stabilize all the complex structure moduli (see, e.g., \cite{Lust:2022mhk, Coudarchet:2023mmm, Lanza:2024uis}).

Discussions so far imply that the typical assumptions for the phenomenological model building, the reliable 4-dimensional effective supergravity framework with all the axio-dilaton and the complex structure moduli stabilized by the fluxes, do not work well in the strict asymptotic regime.
But since the perturbative control is lost in the deep interior  of the field space,  we need to look for an appropriate intermediate regime between the strict asymptotic regime and the deep interior in which  the moduli values are still larger than ${\cal O}(1)$ and the model is well protected from  enhancement of higher order corrections.
For example, in the large volume scenario \cite{Balasubramanian:2005zx}, the K\"ahler moduli are stabilized through the balancing the leading $\alpha'$ correction to the K\"ahler potential against the non-perturbative correction to the superpotential.
For this model to work, other corrections to the potential from the $\alpha'$ (or equivalently, $1/{\cal V}$) as well as the $g_s$ expansions must be sufficiently suppressed.
This has led to studies on the restriction on the values of ${\cal V}$ and $g_s$ consistent with the perturbative control \cite{Junghans:2022exo, ValeixoBento:2023nbv} (for a discussion on the control issue of the large volume scenario in connection with the tadpole conjecture, see, e.g., \cite{Gao:2022fdi}. for application to the cosmological model, see, e.g., \cite{Cicoli:2024bxw}).

The corrections depend on the complex structure moduli as well, but the exact functional forms are not explicitly calculable except for some specific internal manifold geometries, such as those containing the factorized torus \cite{Berg:2005ja, Berg:2005yu}.
For this reason, the complex structure moduli dependent parts  are often treated as  free parameters.
Nonetheless, it might be worth to investigate the behavior of the complex structure moduli in the explicitly calculable models, the feature of which may also appear in other (at least some restricted class of) models.
In particular, we pay our attention to the Sp$(2(h^{2,1}+1))$ ($h^{2,1}$ : the Hodge number of the harmonic $(2, 1)$-form) symmetry of the K\"ahler potential for the complex structure moduli, and   try to find the functional dependence of the corrections to  the K\"ahler potential on  the complex structure moduli allowed by the symmetry, focusing on the leading term for the large field values.
Whereas this symmetry is broken by, for example,  gauge fields or the supersymmetry breaking effects, presumably, depending on the model, the breaking effects may not affect the behavior of the corrections  for the large field values.
In fact,  it is what actually happens in the calculable models in \cite{Berg:2005ja, Berg:2005yu}.
Therefore, we can at least say that there is no principle to suppress the corrections to the K\"ahler potential for the extremely large values of the complex structure moduli, which sets the upper bounds on the field values.

 Our approach in this article is as follows.
 We will compare the behavior of the complex structure moduli with that of the axio-dilaton, since they respect the Sp$(2(h^{2,1}+1))$ and the SL$(2, \mathbb{Z})\cong$ Sp$(2, \mathbb{Z})$ symmetries which have the similar structure as reviewed in Sec. \ref{Sec:symmetry}. 
 In the Einstein frame, the correction to the K\"ahler potential behaves as  $\sim g_s^{-3/2}$ \cite{Becker:2002nn}, which in fact  is a part of a function invariant under the SL$(2, \mathbb{Z})$ symmetry of the axio-dilaton K\"ahler potential \cite{Green:1997tv}.  
 At the same time, ${\cal V}\gg 1$, the condition for the perturbative control,  is rewritten as ${\cal V}_E \gg g_s^{-3/2}$, where ${\cal V}_E$ is the volume in the Einstein frame.
 Whereas it just states that the value of ${\cal V}$ should not be suppressed, we may interpret it in the language of the Einstein frame that the diverging behavior of the correction to the K\"ahler potential in the limit $g_s\to 0$ appears by the SL$(2, \mathbb{Z})$ symmetry but can be prevented by the lower bound on $g_s$, or equivalently, the upper bound on the dilaton field value, to keep  ${\cal V}_E$, hence ${\cal V}$,    sufficiently large.
 Since ${\cal V}$ and $g_s$ are connected to the the KK and the string mass scales, such a bound on the dilaton field value is consistent with the distance conjecture, as discussed in  Sec.  \ref{Sec:dilaton}.
 Motivated by this, we consider  the possibility that the correction to the K\"ahler potential contains the Sp$(2(h^{2,1}+1))$  invariant function of the complex structure moduli  presented in Sec. \ref{Sec:symmetry} which diverges for the infinitely large field value. 
 Indeed, this is what happens in the explicitly calculable models \cite{Berg:2005ja, Berg:2005yu}, suggesting a possibility that  the perturbative control breaks down in the strict asymptotic regime of the complex structure moduli space.
 Moreover, since the correction also depends on ${\cal V}$ and $g_s$ which control the sizes of the KK and the string mass scales, the condition for the perturbative control providing the upper bounds on the  complex structure moduli values can be written in terms of   tower mass scales, as  done in   Sec. \ref{Sec:CSM}.
 As we will see, these bounds look quite similar to what claimed in the distance conjecture.
After discussing the physical implications, we conclude.
The appendix is devoted to the review on the correction to the K\"ahler potential obtained in the calculable models \cite{Berg:2005ja, Berg:2005yu}, which explicitly shows the expected behavior of the complex structure moduli explored in this article.

\section{Symmetry of the K\"ahler potential and invariant function}
\label{Sec:symmetry}

 The main topic of this section is the function of the complex structure moduli (axio-dilaton) which is invariant under the Sp$(2(h^{2,1}+1))$ (SL$(2, \mathbb{Z})$) symmetry appearing in the tree-level K\"ahler potential and becomes dominant in the large field limit, which will be useful in the following discussions.
To investigate this, we begin  with reviews on the structure of the K\"ahler potential for the complex structure moduli and the Sp$(2(h^{2,1}+1))$ symmetry.

We first consider the special K\"ahler geometry arising in the ${\cal N}=2$ supersymmetric couplings of vector multiplets to supergravity.
In particular, we focus on type IIB compactification on a Calabi-Yau 3-fold (CY$_3$) $X_6$, where scalars in the vector multiplets are given by the complex structure moduli. 
Let $(A^I, B_I)$ ($I=0, \cdots, h^{2,1}(X_6)$) be the canonical homology basis of $H_3(X_6, \mathbb{Z})$ with intersection numbers $A^I \cdot A^J=0=B_I\cdot B_J$ and $A^I\cdot B_J=\delta^I_J$.
Then their Poincar\'e dual $(\alpha_I, \beta^I)$ satisfying
\dis{\int_{A^J}\alpha_I=\int_{X_6} \alpha_I\wedge \beta^J =-\int_{B_I}\beta^J=\delta_I^J}
form the basis of $H^3(X_6, \mathbb{Z})$, and the holomorphic $(3,0)$-form can be written as
\dis{\Omega_3=Z^I\alpha_I-{\cal F}_I\beta^I,}
where $Z^I$ and ${\cal F}_I$ correspond to the periods of $\Omega_3$ over cycles $A^I$ and $B_I$, respectively : 
\dis{Z^I=\int_{A^I} \Omega_3,\quad\quad {\cal F}_I=\int_{B_I} \Omega_3.}
Moreover, one can define the prepotential ${\cal F}(Z^I)=(Z^0)^2 F(z^a)$ (in other words, ${\cal F}(\lambda Z^I)=\lambda^2 {\cal F}(Z^I)$) with $z^a=Z^a/Z^I$ ($a=1,\cdots,$ $h^{2,1}(X_6)$) such that ${\cal F}_I =\partial {\cal F}/\partial Z^I\equiv \partial_I{\cal F}$. 
Then the K\"ahler potential for the complex structure moduli is written as
\dis{K_{\rm cs}&=-\log\Big[i\int_{X_6} \Omega_3\wedge\overline{\Omega}_3\Big]
\\
&=-\log[i(\overline{Z}^I{\cal F}_I-Z^I\overline{\cal F}_I)]=-\log\big[|Z^0|^2(2(F-\overline{F})-(z^a-\overline{z}^a)(\partial_a F+\overline{\partial_a F}))\big],\label{eq:KahlerCS}}
after which we can set $Z^0=1$.
From the first expression of the second line, it is clear that the K\"ahler potential is invariant under the Sp$(2(h^{2,1}+1))$ transformation of the periods $\Pi=(Z^I, {\cal F}_I)^{T}$.
That is, since 
\dis{\Sigma =\left(
\begin{array}{cc}
\int_{X_6} \alpha_I\wedge \alpha_J & \int_{X_6} \alpha_I\wedge \beta^J \\
\int_{X_6} \beta^I\wedge \alpha_J & \int_{X_6} \beta^I\wedge \beta^J
\end{array}\right) =
\left(
\begin{array}{cc}
0 & 1 \\
-1 & 0
\end{array}\right)
}
with each block of size $2(h^{2,1}+1) \times 2(h^{2,1}+1)$, the combination $\overline{Z}^I{\cal F}_I-Z^I\overline{\cal F}_I=\overline{\Pi}^T\cdot \Sigma \cdot\Pi$  is invariant under $\Pi \to M\cdot \Pi$, where $M$ is the real matrix satisfying $M^T\cdot \Sigma \cdot M=\Sigma$.

 We can compare this Sp$(2(h^{2,1}+1))$ symmetry  with the SL$(2, \mathbb{Z})$ symmetry of the axio-dilaton $\tau=C_0+ie^{-\Phi}$, also known as the S-duality. 
 The kinetic term for $\tau$,
 \dis{-\frac{1}{2\kappa_{10}^2}\int d^{10}x \sqrt{-G_{10}} \frac{\partial_M\tau\partial^M\overline{\tau}}{2({\rm Im}\tau)^2},\label{eq:taukin}}
 is invariant under the SL$(2, \mathbb{R})$ transformation,
 \dis{\tau \to \frac{a\tau+b}{c\tau+d},\quad\quad ad-bc=1.\label{eq:SL(2)}}
 In type IIB supergravity, it becomes the symmetry of the whole effective action as the 2-form fields also transform as 
 \dis{\left(
\begin{array}{c}
C_2 \\
B_2
\end{array}\right) \to \left(
\begin{array}{cc}
a & b \\
c & d
\end{array}\right)  
\left(
\begin{array}{c}
C_2 \\
B_2
\end{array}\right). }
In the flux compactifications, the fluxes of the 3-form field strengths $F_3=dC_2$ and $H_3=dB_2$ are quantized, then the  SL$(2, \mathbb{R})$ is reduced to  SL$(2, \mathbb{Z})$.
Moreover, from
\dis{\left(
\begin{array}{cc}
a & c \\
b & d
\end{array}\right)
\left(
\begin{array}{cc}
0 & 1 \\
-1 & 0
\end{array}\right)
\left(
\begin{array}{cc}
a & b \\
c & d
\end{array}\right)=
\left(
\begin{array}{cc}
0 & 1 \\
-1 & 0
\end{array}\right),}
it is evident that SL$(2)\cong$ Sp$(2)$, hence the behavior of $\tau$ is expected to be similar to that of the complex structure moduli.
Indeed, $\tau$ may be regarded as the complex structure moduli of CY$_1$, the torus $\mathbb{T}^2$,
\footnote{This fact plays an essential role in the formulation of the F-theory \cite{Vafa:1996xn}, as well as the argument for the connection to the M-theory \cite{Schwarz:1995dk, Aspinwall:1995fw}.} where the periods of the holomorphic $(1,0)$-form $\Omega_1$ over two $1$-cycles $A$ and $B$ are given by
\dis{Z^0=\int_A\Omega_1, \quad\quad Z^0\tau =\int_B\Omega_1, }
respectively.
Then in terms of the Poincar\'e dual $(\alpha, \beta)$, $\Omega_1$ is written as $Z^0(\alpha-\tau \beta)$, which gives the K\"ahler potential consistent with the kinetic term,
\dis{K_\tau =-\log\Big[-i\int_{\mathbb{T}^2} \Omega_1\wedge\overline{\Omega}_1\Big]= -\log\big[-i|Z^0|^2(\tau-\overline{\tau})\big].}
We also note that the SL$(2, \mathbb{Z})$ transformation of $\tau$ can be written in the form of the Sp$(2, \mathbb{Z})$ transformation,
 \dis{\left(
\begin{array}{c}
Z^0 \\
Z^0\tau
\end{array}\right) \to \left(
\begin{array}{cc}
d & c \\
b & a
\end{array}\right)  
\left(
\begin{array}{c}
Z^0 \\
Z^0\tau
\end{array}\right),}
under which $K_\tau$ is invariant before setting  $Z^0=1$.

 The fact that  SL$(2, \mathbb{Z})$   is a special case of  Sp$(2(n+1), \mathbb{Z})$, i.e., $n=0$, implies that so far as the K\"ahler potential is concerned, the behavior of $\tau$ is quite similar to that of the complex structure moduli enjoying the Sp$(2(h^{2,1}+1))$ symmetry.
 In particular, if one imposes that the   quantum corrections to the K\"ahler potential  are invariant under the Sp$(2(h^{2,1}+1))$ and the SL$(2, \mathbb{Z})$ transformations, the K\"ahler potential depends on the complex structure moduli and $\tau$ in the similar  way  even at the quantum level.
 To see this more concretely, we consider the case in which the quantum effects that correct the K\"ahler potential are described by the couplings of the complex structure moduli to the lattice $\Lambda=\{L=(n_I, m^I)^T |n_I, m^I \in \mathbb{Z}, I=0,\cdots,h^{2,1}\}$ representing, for example, the KK/winding modes or instanton effects.
 We can say that the lattice is invariant under the Sp$(2(h^{2,1}+1))$ transformation provided for any $M \in$ Sp$(2(h^{2,1}+1))$, the transformed lattice vector $M^T\cdot L$ belongs to the original lattice $\Lambda$ such that the sum of the combinations $L^T \cdot \Pi =n_I Z^I+m^I{\cal F}_I$ over $n_I$ and $m^I$ is invariant under Sp$(2(h^{2,1}+1))$.
 Then one immediately writes down the Sp$(2(h^{2,1}+1))$ invariant combinations $\sum_\Lambda(\overline{\Pi}^T\cdot \Sigma \cdot \Pi)^s (L^T \cdot \Pi)^p$ as   candidates for the complex structure moduli dependent part of the quantum corrections to the K\"ahler potential.
 Of particular interest is the Sp$(2(h^{2,1}+1))$ invariant function given by
 \dis{E_s(Z^I, \overline{Z}^I)=\sum_{(n_I, m^I)\ne (0,0)}\frac{\big(-\frac{i}{2}\overline{\Pi}^T\cdot \Sigma \cdot \Pi\big)^s}{|L^T \cdot \Pi|^{2s}}
 =\sum_{(n_I, m^I)\ne (0,0)} \frac{\big(-\frac{i}{2}(\overline{Z}^I{\cal F}_I-Z^I\overline{F}_I)\big)^s}{|n_IZ^I+m^I{\cal F}_I|^{2s}}, \label{eq:E(z)}}
 because it has two interesting properties.
 First, as is obvious by rewriting it as 
 \dis{E_s(Z^I, \overline{Z}^I)=\sum_{(n_I, m^I)\ne (0,0)} \frac{\big(-\frac{i}{2}[2(F-\overline{F})-(z^a-\overline{z}^a)(\partial_a F+\overline{\partial_a F})]\big)^s}{|n_0+n_a z^a+m^0(2F-z^a F_a)+m^a F_a |^{2s}},\label{eq:E(z)p}}
$E_s(Z^I, \overline{Z}^I)$ is independent of $Z^0$ once the physical complex structure moduli $z^a$ are defined (hence from now on we will denote it by $E_s(z^a, \overline{z}^a)$).
 That is, whereas we need to `fix' $Z^0$ to some value just like the gauge fixing, the value of $E_s(z^a, \overline{z}^a)$ is not affected at all by the choice of $Z^0$.
 Second, if there is a complex structure modulus $z$ such that $|\overline{\Pi}^T\cdot\Sigma\cdot\Pi| \to \infty$ (hence $K_{\rm cs}\to \infty$) in the limit $|z|\to\infty$, which is in fact typical,  $E_s(z^a, \overline{z}^a)$ also diverges in the same limit.
 The leading term in this large field limit, i.e., in the asymptotic regime, comes from   the $n_0 \ne 0$, $n_a, m^I = 0$ part of the sum.
 Whereas the exact dependence of this leading term on $z^a$ is determined by the functional form of the prepotential $F$, we can say conservatively that $E_s(z^a, \overline{z}^a)$ behaves at least as $\sim |z|^s$ for some complex structure modulus $z$.
 Thus, in the absence of additional principle forbidding this term, the correction to the K\"ahler potential in the ${\cal N}=2$ supersymmetry framework may diverge in the strict asymptotic regime of the complex structure moduli space as it depends on $E_s(z^a, \overline{z}^a)$.
 This indeed what happens in the explicitly calculable models : while it is slightly modified, the leading behavior for the large field value is not affected.
  Third, the lattice structure is quite typical in the presence of a tower of states, such as the KK or the string tower.
 In the case of the explicitly calculable model in Sec. \ref{Sec:CSM}, the KK tower provides the lattice structure.
Then the correction to the K\"ahler potential in the form of $E_s(z^a, \overline{z}^a)$ is generated by integrating out a tower of states.
Indeed, a tower of states also plays a crucial role in the distance conjecture, which considers the behavior of moduli in the large field limit.
Then the diverging behavior of $E_s(z^a, \overline{z}^a)$   in the large field limit can be connected to the distance conjecture, since these two are effects of a tower of states.
In summary, even though $E_s(z^a, \overline{z}^a)$ may not be the only term in  the correction to the K\"ahler potential, it must exist in the presence of a tower of states, and it allows the diverging behavior in the large field limit which can be discussed in light of the distance conjecture. 
\footnote{
  Indeed, it was recently emphasized in \cite{ValeixoBento:2025yhz} that the moduli potential can be dominated by the Casimir energy contribution having the lattice structure.
The form of the potential can be restricted by  (at least the subgroup of) the S-duality   \cite{Chen:2025rkb}.
The correction to the K\"ahler potential in the form of $E_s(z^a, \overline{z}^a)$ may contribute  to such type of the potential. }

  We can also find the function of $\tau$ respecting the SL$(2, \mathbb{Z})=$Sp$(2, \mathbb{Z})$ symmetry and  satisfying   the similar property to $E_s(z^a, \overline{z}^a)$.
 Replacing $Z^I$ and ${\cal F}_I$ by $Z^0$ and $Z^0\tau$, respectively, we obtain 
 \dis{E_s(\tau)=\sum_{(n, m)\ne (0,0)} \frac{\big(-\frac{i}{2}(\overline{Z}^0(Z^0\tau)-Z^0\overline{(Z^0\tau)})\big)^s}{|n Z^0+m Z^0\tau|^{2s}} 
 = \sum_{(n, m)\ne (0,0)} \frac{({\rm Im}\tau)^s}{|n+m  \tau|^{2s}},}
 which is nothing more than the well known Eisenstein series.
  Under the SL$(2, \mathbb{Z})$ transformation given by \eqref{eq:SL(2)} (with $a, b, c, d \in \mathbb{Z}$), the denominator of the summand becomes $|n'+m'\tau|^{2s}$ with $n'=d n+ b m$ and $m'=a m+c n$, and since the transformed lattice $\{(n', m')\}$ is nothing more than the original lattice, $E_s(\tau)$ is SL$(2, \mathbb{Z})$ invariant.
  Moreover, the second property of $E_s(z^a, \overline{z}^a)$ in the previous paragraph can be found as well : the $n \ne 0$, $m=0$ part of the sum is given by
  \dis{\sum_{m\in \mathbb{Z}-\{0\}}\frac{({\rm Im}\tau)^s}{n^{2s} }=2 ({\rm Im}\tau)^s \sum_{n \in\mathbb{N}}\frac{1}{n^{2s}}=2 ({\rm Im}\tau)^s \zeta(2s), \label{eq:subsum}}
  which behaves as $\sim({\rm Im}\tau)^s$ in the limit ${\rm Im}\tau\to\infty$.

 \section{Dilaton value for perturbative control}
 \label{Sec:dilaton}
 
  Whereas we are interested in the behavior of the complex structure moduli in the corrected K\"ahler potential, we first consider that of the axio-dilaton in this section.
  This is because we learn in Sec. \ref{Sec:symmetry} that the Sp$(2(h^{2,1}+1))$ symmetry obeyed by the complex structure moduli has the similar structure to the SL$(2, \mathbb{Z})$ symmetry of the axio-dilaton, and the latter is much simpler.
  This enables us to understand the diverging behavior of the complex structure moduli which appears in the explicitly calculable model easier from that of the axio-dilaton.
   In particular, in the Einstein frame, it seems that the correction to the K\"ahler potential depends on the  SL$(2, \mathbb{Z})$ invariant Eisenstein series, hence diverges for the large dilaton field value.
 Then for the perturbative control, the dilaton field value is bounded from above, and as well known, this condition is nothing more than the condition ${\cal V}\gg 1$ for the perturbative control in the string frame.

\subsection{Axio-dilaton dependence of the correction to the K\"ahler potential}

It is already known that the metric for the K\"ahler moduli receives the perturbative ${\cal O}({\alpha'}^3)$ corrections generating $R^4$ terms, where $R$ represents the 10-dimensional Riemann tensor \cite{Antoniadis:1997eg}.
This originates from the 4-loop corrections to the world-sheet $\beta$-function in the  sigma model  \cite{Gross:1986iv}.
As a result, the 4-dimensional dilaton is modified by \cite{Becker:2002nn}
\dis{e^{-2\Phi_4} = e^{-2\Phi} {\cal V}  \to e^{-2\Phi} \Big({\cal V}  +\frac12\xi\Big)}
in the string frame, where ${\cal V}$ is the volume of the  CY$_3$ $X_6$ (in units of the string length) in the string frame and
\dis{\xi=-\frac{\chi(X_6)\zeta(3)}{2(2\pi)^3}.}
Here $\chi(X_6)$ is the Euler characteristic of  $X_6$ and $\zeta(3)\simeq 1.202$ is Ap\'ery's constant.
Since   the 10-dimensional metric in the string frame ($G_{MN}^{(s)}$) and that in the Einstein frame ($G_{MN}^{(E)}$) are related by $G_{MN}^{(E)}=e^{-\Phi/2}G_{MN}^{(s)}$,  the 4-dimensional dilaton becomes
\dis{e^{-2\Phi_4} = e^{-2\Phi} \Big(e^{6\times (\Phi/4)}{\cal V}_E +\frac12\xi\Big)= e^{-\Phi/2} \Big( {\cal V}_E +\frac{\xi}{2}e^{-\frac32\Phi}\Big)}
in the Einstein frame.
Therefore, the K\"ahler potential for the K\"ahler moduli is corrected as \cite{Becker:2002nn}
\dis{K_{\rm K}=-2\log\Big[{\cal V}_E+\frac{\xi}{2 g_s^{3/2}}\Big].\label{eq:Kahler1}} 
  The addition of the $\xi$ term breaks the no-scale structure such that the potential for the K\"ahler moduli is modified by
\dis{\delta V=3 \hat{\xi}e^K \frac{(\hat{\xi}^2+7\hat{\xi}{\cal V}_E+{\cal V}_E^2)}{({\cal V}_E-\hat{\xi})(2{\cal V}_E+\hat{\xi})^2} |W_0|^2 \simeq \frac34 \hat{\xi}|W_0|^2\frac{1}{{\cal V}_E^3},}
where $\hat{\xi}=\xi/g_s^{3/2}$ and $W_0$ is the superpotential before the non-perturbative correction.
Whereas it alone does not stabilize the K\"ahler moduli, in the limit ${\cal V}_E \to \infty$, it can compete with the non-perturbative term of the superpotential provided the latter is enhanced by the cycle whose size is exponentially smaller than ${\cal V}_E$.
Then K\"ahler moduli can be stabilized, which is nothing more than the way that  the large volume scenario works.

 On the other hand, discussion  in Sec. \ref{Sec:symmetry} implies that imposing  the SL$(2, \mathbb{Z})$ symmetry of $\tau$ on the corrected K\"ahler potential, the factor $\zeta(3)/g_s^{3/2}$, or equivalently, $\zeta(3)({\rm Im}\tau)^{3/2}$, can be regarded as a dominant part of the SL$(2, \mathbb{Z})$ invariant function  
  \dis{\frac12 E_{3/2} (\tau, \overline{\tau})= \frac12\sum_{(n, m)\ne (0,0)}\frac{({\rm Im}\tau)^{3/2}}{|n+ m\tau|^3}\label{eq:taucorr}}
 in the limit $g_s \to 0$ (hence ${\rm Im}\tau\to\infty$).
 In fact, this was already pointed out in \cite{Green:1997tv}, which conjectured that \eqref{eq:taucorr} can be generated by   the D-brane instanton effects.
 This can be argued in the following way \cite{Green:1997tv}.
 Isolating the $n \ne 0$, $m=0$ part (see \eqref{eq:subsum}), \eqref{eq:taucorr} can be rewritten as
 \dis{\frac12 E_{3/2} (\tau, \overline{\tau})= \zeta(3)({\rm Im}\tau)^{3/2}+\frac{({\rm Im}\tau)^{3/2}}{2}\sum_{m\ne 0, n}\int_0^\infty dy y^{1/2} e^{-y (n+m\tau)(n+m\overline{\tau})}.}
 Using the Poisson resummation formula,
 \dis{\sum_{p=-\infty}^\infty e^{-\pi A(p+x)^2} =\frac{1}{\sqrt{A}}\sum_{m=-\infty}^\infty e^{-\pi\frac{m^2}{A}+2\pi m x},}
 and isolating the $m\ne 0, n=0$ part in addition, it becomes
 \dis{\frac12 E_{3/2} (\tau, \overline{\tau})=& \zeta(3)({\rm Im}\tau)^{3/2}+\frac{\pi^{5/2}}{6 ({\rm Im}\tau)^{1/2}} 
 \\
 &+({\rm Im}\tau)^{3/2}\frac{\pi^{1/2}}{2}\sum_{(n,m)\ne (0,0)}\int dy e^{-\pi^2\frac{n^2}{y}+2\pi i m n {\rm Re}\tau-y m^2{\rm Im}\tau^2}.}
 From the steepest descent method, one finds that the integral over $y$ is dominated by the value of $y$ satisfying $\frac{d}{dy} (-\pi^2\frac{n^2}{y}-y m^2{\rm Im}\tau^2)=0$, i.e., $y=\pm (n/m)\pi/{\rm Im}\tau$, then the integrand can be estimated by exp$[2\pi i m n \tau]$ or exp$[-2\pi i m n \overline{\tau}]$, which may be interpreted as effects of the D-brane instanton and anti-instanton, respectively.

\subsection{Value of the dilaton   consistent with perturbative control}

 In the Einstein frame,  the condition for the perturbative control ${\cal V}\gg 1$ can be rewritten as 
 \dis{{\cal V}_E g_s^{3/2} \gg 1 \label{eq:dilcond}.}
  In other words, the value of ${\cal V}_E$ must be sufficiently sizeable compared to $g_s^{-3/2}$ to keep ${\cal V}\gg 1$.
 Referring to the corrected K\"ahler potential $K_K$ given by \eqref{eq:Kahler1}, this can be interpreted in the Einstein frame that the perturbative control can be lost in the weak coupling limit unless the correction $\xi/(2g_s^{3/2})$ is suppressed compared to ${\cal V}_E$.
This suggests that for the perturbative control, $g_s$ cannot be arbitrarily small, or equivalently, the field value of Im$\tau$ cannot be arbitrarily large.
 \footnote{Since the squared gauge coupling in type IIB string model is proportional to $g_s$, the lower bound on $g_s$ indicates that of the gauge coupling, which reminds us of the weak gravity conjecture \cite{Arkani-Hamed:2006emk, Harlow:2022ich}. }
 Moreover, given fixed value of $M_{\rm Pl}$, the sizes of two tower mass scales, the KK mass scale $M_{\rm KK}$   and the string mass scale $M_s$,  are controlled by ${\cal V}$ as well as $g_s$  so we expect that the condition \eqref{eq:dilcond} can be rewritten as the upper bound on the field value of Im$\tau$ given by the tower mass scales.

   To see this, we define the representative KK scale $M_{\rm KK}$ by parametrizing ${\cal V}$,  the  overall volume of the internal manifold in units of the string length, by ${\cal V}=(M_s/M_{\rm KK})^6$.
 The physical meaning of $M_{\rm KK}$ is clear by noting that ${\cal V}$  is determined by the  stabilized values of the K\"ahler moduli.   
  For example, in type IIB compactification, the K\"ahler form can be expanded in terms of the basis of the harmonic $(1,1)-$form $b^\alpha_{i\overline{j}}$ as $\omega_{i\overline{j}} = \sum_\alpha t_\alpha b^\alpha_{i\overline{j}} $ ($\alpha=1, \cdots h^{1,1}$ ).
  Then the stabilized values of the K\"ahler moduli $t_\alpha$   determine  the size of the 2-cycles of the internal manifold, hence the overall volume as ${\cal V}=(1/6)\kappa^{\alpha\beta\gamma}t_\alpha t_\beta t_\gamma$ where $\kappa^{\alpha\beta\gamma}$ is an intersection number. 
   If we assume the isotropic compactification, i.e., in the absence of the  specific direction   much larger than  other directions in size, $t_\alpha$ are stabilized at the similar value, thus the actual KK mass scales associated with the 2-cycles given by $(M_{\rm KK}^{(\alpha)})^2=t_\alpha M_s^2$ are similar in size.
 In this case,   the representative KK mass scale $M_{\rm KK}$ will more or less coincide with $M_{\rm KK}^{(\alpha)}$.
 If the compactification is anisotropic, $M_{\rm KK}$ does not necessarily coincide with the actual KK mass scale   $M_{\rm KK}^{(\alpha)}$.
Even in this case, the scaling of ${\cal V}$ (hence $M_{\rm KK}$)   leads to the common scaling of $t^\alpha$ (hence $M_{\rm KK}^{(\alpha)}$).
For example, if some parameter scales as ${\cal V}^r$ with some rational number $r$, it scales as $t_\alpha^r$ for every $\alpha$.
It means that every  $M_{\rm KK}^{(\alpha)}$ scales with the same rate as $M_{\rm KK}$.
Then our discussion on the scaling of $M_{\rm KK}$ applies   to the anisotropic case as well.

 Combining the relation ${\cal V}=(M_s/M_{\rm KK})^6$ with \eqref{eq:Ms} we obtain  
  \dis{M_{\rm KK}\sim \frac{M_s}{{\cal V}^{1/6}}\sim \frac{g_s}{{\cal V}^{2/3}}M_{\rm Pl},}
  or equivalently,
  \dis{M_{\rm KK}\sim \frac{M_s}{g_s^{1/4}{\cal V}_E^{1/6}}\sim \frac{M_{\rm Pl}}{{\cal V}_E^{2/3}}\label{eq:KKscale}}
from which the condition \eqref{eq:dilcond} reads
  \dis{{\rm Im}\tau =\frac{1}{g_s} \ll {\cal V}_E^{2/3}\sim   \frac{M_{\rm Pl}}{M_{\rm KK}}.\label{eq:Inqtau}}
  Meanwhile,  the relation $g_s=e^\Phi$ indicates that the weak coupling limit $g_s\to 0$ is equivalent to $\Phi\to -\infty$.
  Indeed, so far as $g_s$ is restricted to be smaller than $1$ for perturbativity, the value of $\Phi$ is negative.
 In this case, one can rewrite \eqref{eq:Inqtau} in terms of $\Phi$ as
   \dis{&|\Phi| \ll   \log\Big(\frac{M_{\rm Pl}}{M_{\rm KK}}\Big),
\quad\quad
  {\rm or}\quad\quad M_{\rm KK}\ll M_{\rm Pl}e^{- |\Phi|}.\label{eq:gsMKK}}  
  This expression is consistent with the distance conjecture \cite{Ooguri:2006in}, which claims that near the asymptotic regime the tower mass scale can be written as $m_t= M_{\rm Pl}e^{-\alpha \varphi}$ for some positive $\alpha$, where  $\varphi$ is the geodesic distance of some modulus.
  Indeed, from the kinetic term for $\tau$ given by \eqref{eq:taukin}, one finds that if the potential for   Im$\tau$ is ignored, the geodesic distance for Im$\tau$ is nothing more than the field value of  $\Phi$.
  Of course, the moduli responsible for the size of $M_{\rm KK}$ are the K\"ahler moduli thus the distance conjecture for $M_{\rm KK}$ is written in terms of the K\"ahler moduli and moreover, in the form of the equality, contrary to the inequality \eqref{eq:gsMKK}.
Nevertheless, the interpretation of \eqref{eq:gsMKK} can be quite similar to the distance conjecture :  since the scale $M_{\rm KK}$ must be high enough to decouple from the 4-dimensional EFT, the value of $\Phi$ cannot be too large to keep $M_{\rm Pl}e^{-2|\Phi|}$ to be larger than $M_{\rm KK}$.
Otherwise, the perturbative control will be lost as the string frame volume ${\cal V}$ is not sufficiently large, thus the EFT ignoring the correction $\xi/(2g_s^{3/2})$ breaks down.

  The relation  \eqref{eq:KKscale} also can be used to obtain the upper bound on the string mass scale, $M_s \sim g_s^{1/4}(M_{\rm Pl}/{\cal V}_E^{1/2})$ : 
  \dis{\frac{1}{g_s} \ll {\cal V}_E^{2/3}\sim g_s^{1/3}\Big(\frac{M_{\rm Pl}}{M_s}\Big)^{4/3},}
  which can be rewritten as
  \dis{{\rm Im}\tau =\frac{1}{g_s} \ll  \frac{M_{\rm Pl}}{M_s}.}
In the distance conjecture-like form, it reads 
 \dis{&|\Phi| \ll  \log  \Big(\frac{M_{\rm Pl}}{M_s}\Big),
 \quad\quad
  {\rm or}\quad\quad  M_s\ll M_{\rm Pl}e^{- |\Phi|}.\label{eq:gsMS}}
  This sets the upper bound on the field value of $|\Phi|$  for the EFT to decouple from  the string excitations. 
  
 The bounds we considered so far also can be interpreted that for consistency with the perturbative  control, the tower mass scales $M_{\rm KK}$ and $M_s$ are required to be suppressed by at least $e^{-\alpha|\Phi|}$ for some positive $\alpha\sim {\cal O}(1)$ compared to $M_{\rm Pl}$.
 From these, we expect that the species scale above which quantum gravity effects are no longer negligible is bounded from above \cite{Veneziano:2001ah, Dvali:2007hz, Dvali:2007wp, Dvali:2009ks, Dvali:2010vm, Dvali:2012uq}.
To see this, we first note that given the tower mass scale $m_t$ with respect to  which the spectrum of states in a tower can be written as $m_n=n^{1/p}m_t$, the species number $N_{\rm sp}$ associated with the tower is defined by  the number of tower  sates with mass below the species scale $\Lambda_{\rm sp}$.
This is estimated as $N_{\rm sp}=(\Lambda_{\rm sp}/m_t)^p$.
Since the species scale  is given by $\Lambda_{\rm sp}=M_{\rm Pl}/\sqrt{N_{\rm sp}}$,
\footnote{This is evident from, for example, the fact that the 1-loop correction to the graviton propagator generated by $N_{\rm sp}$ states is proportional to  $N_{\rm sp}{p^2}/{M_{\rm Pl}^2}$.
This is suppressed compared to the tree-level propagator provided $|p|<\Lambda_{\rm sp}=M_{\rm Pl}/\sqrt{N_{\rm sp}}$.
}
 we obtain \cite{Castellano:2021mmx}
 \dis{\Lambda_{\rm sp}= M_{\rm Pl}^{\frac{2}{p+2}}m_t^{\frac{p}{p+2}},\quad\quad N_{\rm sp}=\Big(\frac{M_{\rm pl}}{m_t}\Big)^{\frac{2p}{p+2}}.\label{eq:spKK}}
 In the case of the KK tower, $m_t=M_{\rm KK}$ and $p$ corresponds to the number of KK towers of the same mass scale.
  When the $6$ extra dimensions are similar in size, $p$ is just given by $6$, hence the species scale/number associated with the KK tower become
 \dis{\Lambda^{\rm (KK)}_{\rm sp}= M_{\rm Pl}^{1/4}M_{\rm KK}^{3/4},\quad\quad N^{\rm (KK)}_{\rm sp}=\Big(\frac{M_{\rm pl}}{M_{\rm KK}}\Big)^{3/2},}
 respectively.
 Then the bound \eqref{eq:gsMKK} provides the upper(lower) bound on $\Lambda^{\rm (KK)}_{\rm sp}$($N^{\rm (KK)}_{\rm sp}$) :
 \dis{\Lambda^{\rm (KK)}_{\rm sp}\ll M_{\rm Pl}e^{-(3/4)|\Phi|},\quad\quad N^{\rm (KK)}_{\rm sp} \gg e^{(3/2)|\Phi|}=\frac{1}{g_s^{3/2}}.}
 On the other hand, for the string tower, $m_t=M_s$ and $m_n\sim \sqrt{n}M_s$ but the exponentially growing degeneracy $d_n \sim e^{\sqrt{n}}$ \cite{Kani:1989im} implies that $p\to \infty$ limit has to be taken \cite{Castellano:2021mmx}.   The $p\to \infty$ limit for the string tower can be intuitively understood from the facts that the degeneracy grows exponentially faster than any other polynomial and that the species scale is given by more fundamental scale $M_s$, just like the case of the KK tower where the species scale in \eqref{eq:spKK} is nothing more than the higher dimensional Planck scale (which is also more fundamental scale).
  This also coincides with the behavior of the tensionless string considered in the emergent string conjecture \cite{Lee:2019wij}. 
 Then we obtain the simple relations
 \dis{\Lambda_{\rm sp}^{(s)}=M_s,\quad\quad N_{\rm sp}^{(s)}=\Big(\frac{M_{\rm pl}}{M_s}\Big)^2,\label{eq:spS}}
 and regarding \eqref{eq:gsMKK}, they are bounded as
 \dis{\Lambda_{\rm sp}^{(s)} \ll  M_{\rm Pl}e^{- |\Phi|},\quad\quad N_{\rm sp}^{(s)} \gg e^{2|\Phi|}=\frac{1}{g_s^2}.}

 \section{Complex structure moduli value for perturbative control}
\label{Sec:CSM}

\subsection{Complex structure moduli dependence of the correction to the K\"ahler potential}
 
 Discussions on the axio-dilaton $\tau$ in Sec. \ref{Sec:dilaton} show that in the Einstein frame, the correction to the K\"ahler potential depends on the SL$(2, \mathbb{Z})$ invariant function of $\tau$, which allows the diverging behavior in the weak coupling limit Im$\tau\to\infty$.
 This sets the  lower (upper) bound on $g_s$ (Im$\tau$), and it is nothing more than the condition for the perturbative control by the large string frame volume ${\cal V}$.
 Meanwhile, as pointed out in Sec. \ref{Sec:symmetry}, the SL$(2, \mathbb{Z})$ symmetry of $\tau$ and the Sp$(2(h^{2,1}+1))$ symmetry of the complex structure moduli have the similar structure.
From these, we expect that when the contribution of the complex structure moduli to the correction to the K\"ahler potential is regulated by the Sp$(2(h^{2,1}+1))$ symmetry, the large volume limit may not prevent the large values of the complex structure moduli, then the corrections to the K\"ahler potential are not suppressed.
This sets the upper bound on the values of the complex structure moduli for the reliable perturbative expansion.

 Concerning  type IIB orientifold compactification with D$3$- and D$7$-branes which is useful for phenomenological applications, the string loop corrections at ${\cal O}(g_s^2{\alpha'}^2)$ and ${\cal O}(g_s^2{\alpha'}^4)$ (in the string frame) or ${\cal O}(g_s{\cal V}_E^{-2/3})$ and ${\cal O}({\cal V}_E^{-4/3})$ (in the Einstein frame) are given by \cite{Berg:2005ja, Berg:2005yu} (see also \cite{Berg:2007wt, Cicoli:2007xp} for an intuitive interpretation)
 \dis{\delta K_{\rm KK}=g_s\sum_i c_i^{\rm KK}(z^a, \overline{z}^a)\frac{t_i^\perp}{{\cal V}_E},\quad\quad 
 \delta K_{\rm W}= \sum_j c_j^{\rm W}(z^a, \overline{z}^a)\frac{1}{t_j^\cap {\cal V}_E},\label{eq:deltaKs}}
 respectively.
 For $\delta K_{\rm KK}$, the index $i$ runs over the pairs of parallel D-branes/O-planes and $t_i^\perp$ indicates the K\"ahler modulus for the 2-cycle perpendicular to the $i$th pair of parallel branes.
 As can be inferred from $t_i^\perp \sim 1/M_{\rm KK}^2$, $\delta K_{\rm KK}$ is generated by the KK strings stretching between parallel stacks of branes. 
  For $\delta K_{\rm W}$, which is generated by   strings   winding the intersection 2-cycles   between   stacks of D-branes/O-planes, the index $j$ runs over the pairs of intersecting branes and $t_j^\cap$ indicates the K\"ahler modulus for the intersection 2-cycle of the $j$th pair of  branes.
  \footnote{Of course, the winding modes appear in the toroidal compactificaiton : for the Calabi-Yau manifold,  the fact that $h^{1, 0}=0$ indicates that there is no 1-cycle on which the string wrapping, hence the winding modes cannot appear. \label{foot:CY} }
  Meanwhile, $c_i^{\rm KK}(z^a, \overline{z}^a)$ and $c_j^{\rm W}(z^a, \overline{z}^a)$ are functions of the complex structure moduli.
  Whereas their exact expressions are unknown except for some specific examples, if we impose the Sp$(2(h^{2,1}+1))$ symmetry, it may not be strange that they contain $E_s (z^a, \overline{z}^a)$ thus diverge in the limit $|z^a|\to \infty$,   threatening the perturbative control.
 In this case,  the values of the complex structure moduli must be bounded from above for consistency with the perturbative control.
  Indeed,   in the calculable examples   in \cite{Berg:2005ja, Berg:2005yu} where the single complex structure modulus $U$ for the $\mathbb{T}^2$ factor of the internal manifold (along, say, $4, 5$ directions) does not mix with other complex structure moduli, the tree-level K\"ahler potential is invariant under the Sp$(2)=$SL$(2)$ transformation of $U$ in the absence of the gauge fields associated with the brane.
  Moreover, even if the symmetry breaking gauge fields represented by $A$ are taken into account,   $\delta K_{\rm KK}$ and $\delta K_{\rm W}$ contain a function $E_2(U, \overline{U} ; A)$ which  has the similar structure to $E_2(U, \overline{U})$.
  More explicitly,  
  \dis{E_2(U, \overline{U} ; A)=\sum_{(n^4, n^5)\ne (0,0)}e^{2\pi i n^p a_p}\frac{U_2^2}{|n^4+Un^5|^4},\label{eq:EUUA}}
 where  $a_p$ ($p=4, 5$) indicates the gauge field along $\mathbb{T}^2$, from which we define $A=U a_4-a_5$.
 Then one immediately finds that $E_2(U, \overline{U} ; A)$ becomes $E_2(U, \overline{U})$ when $a_p=0$, or equivalently, $A=0$.
 \footnote{Under $U \to (aU+b)/(cU+d)$, $a_4$ and $a_5$ in $E_2(U, \overline{U}; A)$ become  $\tilde{a}_4=a a_4-c a_5$ and $\tilde{a}_5=-b a_4+d a_5$, respectively, which shows that  $E_2(U, \overline{U}; A)$ is not invariant under the SL$(2)$ transformation (see also discussion below \eqref{eq:loopE2}).
 Meanwhile, one may introduce the fictitious SL$(2)$ transformations $a_4\to d a_4+c a_5$ and $a_5 \to  b a_4+a a_5$, such that $\tilde{a}_{4, 5}=a_{4, 5}$ hence $E_2(U, \overline{U}; A)$ becomes SL$(2)$ invariant. }
 We need to note here that even in the presence of the gauge field $A$, the leading behavior of $E_2(U, \overline{U} ; A)$ in the limit $U\to\infty$ is the same as that of $E_2(U, \overline{U})$.
 Indeed, $E_2(U, \overline{U} ; A)$ can be expanded as
 \dis{  E_2(U, \overline{U};A)=&2\pi^4\Big(\frac{U_2^2}{90}-\frac13  A_2^2+\frac23 \frac{A_2^3}{U_2}-\frac13\frac{A_2^4}{U_2^2}\Big)
\\
&+ \frac{\pi}{2 U_2} \big[{\rm Li}_3(e^{2\pi i A})+2\pi A_2{\rm Li}_3(e^{2\pi i A}) +{\rm c.c.}\big]
\\
&+ \frac{\pi^2}{U_2}\sum_{m>0}\big[(mU_2-A_2){\rm Li}_2(e^{2\pi i (mU-A)})+ (mU_2+A_2){\rm Li}_2(e^{2\pi i (mU+A)}))
\\
&\quad\quad\quad\quad\quad +{\rm c.c.}\big]
\\
&+ \frac{\pi}{2 U_2}\sum_{m>0}\big[ {\rm Li}_3(e^{2\pi i (mU-A)})+{\rm Li}_3(e^{2\pi i (mU+A)})+{\rm c.c.}\big],\label{eq:E2zz}}
where  $U_2$ and $A_2$ denote the imaginary parts of $U$ and $A$, respectively.
From this, one realizes that in the limit $U_2\to\infty$, $ E_2(U, \overline{U};A)$ is dominated by the $U_2^2$ term, just like $E_2(U, \overline{U})$.

 It may be worth to investigate how the string 1-loop correction containing $E_2(U, \overline{U} ; A)$ is obtained   in the explicitly calculable examples.
 For this purpose, we note that the corrections to the K\"ahler potential \eqref{eq:deltaKs} can be read off  from the 2-point correlation function of the vertex operators for the moduli.
 The 1-loop correction to this is given by the integral of the 2-point correlation function of the world-sheet field  over the modular parameter for the compact string world-sheet topology of genus one.
 However, the calculation described above is lengthy and not an original work of this article, so we sketch it in App. \ref{App:example} and summarize the main points below :
  \begin{itemize}
  \item In the string 1-loop calculation, $E_2(U, \overline{U} ; A)$ appears quite naturally.
   The world-sheet (Euclidean) action $S_E$ in the integral weight $e^{-S_E}$ contains  $|n^4+n^5 U|^2/U_2$ ($n^4, n^5\in\mathbb{Z}$)  which originates from the structure of the metric for the internal manifold, and the 2-point correlation function of the world-sheet scalar is given by the linear polynomial in  this combination. 
 In the integral over the modular parameter, this structure plays the crucial role to obtain  $E_{2}(U, \overline{U} ;A)$.  
  The dependence on the gauge field $A$ comes from the boundary world-sheet action.
  \item As we have seen, the function $E_2(U, \overline{U} ; A)$ becomes the SL$(2)$ invariant function $E_2(U, \overline{U})$ for the vanishing gauge field ($A=0$).
  The SL$(2)$ breaking effect by the gauge fields is consistent with the symmetry breaking structure of the tree-level kinetic term for $A$.
  Moreover, since the gauge fields appear in the string 1-loop calculation through the boundary world-sheet action, the imaginary part of $S_E$ (see \eqref{eq:SB}) in the weight $e^{-S_E}$,  $E_2(U, \overline{U} ; A)$ depends on the gauge fields through the phase (and in the form of the coupling to the lattice) in the summand.
  \footnote{Since $a_4=A_2/U_2$, the phase $2\pi i(n^4 a_4+n^5 a_5)$ in the $n_4\ne 0, n_5=0$ part of the sum (which gives the dominant term of $E_2(U, \overline{U} ; A)$) in \eqref{eq:EUUA} becomes zero in the $U_2\to \infty$ limit for a finite value of $A_2$.
  This structure is guaranteed by the consistency with the symmetry breaking by the kinetic term for the gauge field.
  We may generalize this by, for example, defining   the function $(\overline{\Pi}^T\cdot \Sigma\cdot  Q-\Pi^T\cdot\Sigma\cdot \overline{Q})/(\overline{\Pi}^T\cdot \Sigma\cdot\Pi)$ where $Q=Z_0(a_4, a_5)^T$, which in fact is $a_4$ and manifestly shows that it  vanishes as $|\overline{\Pi}^T\cdot \Sigma\cdot\Pi|\to\infty$.
  As can be found in the D9-brane action in \eqref{eq:L}, the combination $\overline{\Pi}^T\cdot \Sigma\cdot  dQ$ is exactly what appears in the kinetic term for the gauge fields.
  In  the $h^{2,1}$ complex structure moduli case, if a function of the same structure couples to $n_0$ in \eqref{eq:E(z)p} in the phase, the phase in the $n_0\ne 0, n_a, m^I=0$ part of the sum vanishes in the limit $\overline{\Pi}^T\cdot \Sigma\cdot\Pi \to\infty$, hence the sum is the same as that  in the absence of the gauge fields.
  \label{foot:case}}
 Since this phase in the  $n_4\ne 0, n_5=0$ part of the sum (which gives the dominant term of $E_2(U, \overline{U} ; A)$) vanishes in the limit $|U| \to \infty$, the behavior of $E_2(U, \overline{U} ; A)$ is the same as $E_2(U, \overline{U})$ in the same limit.
  \item Whereas $c_i^{\rm KK}(U, \overline{U})$ and   $c_j^{\rm W}(U, \overline{U})$ are given by linear combinations of $E_2(U, \overline{U} ; A)$ with various forms of $A$, there is no cancellation of the leading terms $\sim U_2^2$ in the limit $|U| \to \infty$.
  Therefore, both $c_i^{\rm KK}(U, \overline{U})$ and   $c_j^{\rm W}(U, \overline{U})$ behave as $\sim U_2^2$ in this limit.
  \end{itemize}
Even if we consider the multiple, i.e., $h^{2,1}$, complex structure moduli, we expect these three properties still hold.
The Sp$(2(h^{2,1}+1))$ invariance of the complex structure moduli will be encoded in the metric for the internal manifold $G_{mn}$, hence in the world-sheet action $S_{P} \sim \int d^2w G_{mn} \partial X^m \overline{\partial} X^n$.
 Combined with the KK modes in $\partial X^m$, the structure like $|L^T\cdot\Pi|^2/(\overline{\Pi}^T\cdot\Sigma\cdot \Pi)$ can appear in $S_P$.
  Even if the symmetry is broken, the breaking effect can be controlled by, for example, the gauge fields in the boundary action.
  Indeed,  the gauge field dependence appears from the    coupling between the lattice (the KK modes in our example) and the gauge fields along the internal directions.
 In the Euclidean path integral formalism, it is an open string contribution, which appears as a phase  (see \eqref{eq:SB} and also \eqref{eq:EUUA}), generating the interference between  terms in the summation \eqref{eq:EUUA}.
 However, the closed string contribution (which is irrelevant to the gauge field at tree level hence respects the  Sp$(2(h^{2,1}+1))$ symmetry) gives, after the Gaussian integral, the structure that each term in \eqref{eq:EUUA} diverges due to the contribution $n^4=0$.
 Then we may say that so far as the closed string contribution can be well separated from the open string contribution, which is typical when the higher genus contributions are suppressed, the resulting  correction to the K\"ahler potential preserves the diverging behavior in the limit of the large  complex structure moduli values.  
   Away from such microscopic structure, we may argue that so far as the correction to the K\"ahler potential is a smooth function of the gauge field, as can be seen in the behavior of $E_2(U, \overline{U};A)$, there exists the direction in the field space where at least one of complex structure moduli takes the value much larger than the gauge field value. 
 This corresponds to the limit where the  Sp$(2(h^{2,1}+1))$ symmetry is restored, and  our discussion applies to this case.
 Hence, while it is difficult to verify that the large complex structure behavior is always protected from the correction by the gauge fields, we may say that this is quite generic.
 Of course, besides the issue of generality, the existence of the model showing the  large complex structure behavior discussed above is worth to investigate, as the perturbative control is important for the validity of the model.

  Therefore, motivated by the calculable examples (see footnote \ref{foot:case}), we focus on the case where the large field behaviors of $c_i^{\rm KK}(z^a, \overline{z}^a)$ and $c_j^{\rm W}(z^a, \overline{z}^a)$ are not  affected by the value of $A$ which is realized when $A$ appears in the phase   given by the coupling to the lattice and  vanishes in the limit $|U|\to\infty$.
  Then we   conjecture that the factors $c_i^{\rm KK}(z^a, \overline{z}^a)$ and $c_j^{\rm W}(z^a, \overline{z}^a)$ in the loop corrections $\delta K_{\rm KK}$ and $\delta K_{\rm W}$ contain  the term denoted by $E_s(z^a, \overline{z}^a ; A)$, which is reduced to \eqref{eq:E(z)} for $A=0$ and diverges as $|z|^{s}$ for some complex structure modulus in the limit $|z|\to\infty$  : 
 \dis{\delta K_{\rm KK} \sim g_s E_s(z^a, \overline{z}^a ; A) \frac{t_i^\perp}{{\cal V}_E},\quad\quad 
 \delta K_{\rm W}\sim E_s(z^a, \overline{z}^a ; A) \frac{1}{t_j^\cap {\cal V}_E}. \label{eq:CSMCon} }
 The examples in \cite{Berg:2005ja, Berg:2005yu} correspond to the case of $s=2$.
  The numerical coefficients are model dependent,   determined by the loop factor and the number of redundancies giving the same functional dependence on moduli.
 In the calculable example, the loop factor is given by $1/(256\pi^6)$ and the number of redundancies does not exceed ${\cal O}(10^2)$.
 So we expect that except for the  case where some specific correction must be forbidden (for example, in Calabi-Yau manifold, $\delta K_W=0$ : see footnote \ref{foot:CY})  the numerical coefficients are given by the typical loop factor we encounter in the field theory calculations.
 Indeed, in the next subsection, we will take the logarithm of the above expression, in which case   the size of the coefficient does not  play the crucial role.

\subsection{Values of the complex structure moduli   consistent with perturbative control}

 When the string loop correction to the K\"ahler potential  is given in the form of \eqref{eq:CSMCon}, it becomes very large in the asymptotic regime of the complex structure moduli space, resulting in the breakdown of the perturbative control.
 This comes from the expected behavior of $E_s (z^a, \overline{z}^a ; A)$ which grows at least as $|z|^s$ for some complex structure modulus $z$ in the limit $|z| \to\infty$. 
 This is what we can find in the explicitly calculable models, where $z=U$ and $s=2$.
 Therefore, for the perturbative control, the value  of $z$ cannot be too large, but restricted by the values of ${\cal V}$ and $g_s$.
 From this, we can connect the upper bound on the value  of $z$ to the tower mass scales $M_{\rm KK}$ and $M_s$ in the distance conjecture-like form.
 
  To see this, we note that since the tree-level K\"ahler potential for the K\"ahler moduli is given by $-2\log{\cal V}_E$, the correction to the K\"ahler potential $\delta K$ can be regarded as the ratio $\delta {\cal V}_E/{\cal V}_E$  :
 \dis{-2\log{\cal V}_E+\delta K \simeq -2\log\Big({\cal V}_E-\frac{{\cal V}_E}{2}\delta K\Big),}
 hence the perturbativity condition reads $\delta K\ll 1$.
 For $\delta K_{\rm KK}$, the condition $\delta K_{\rm KK} \ll 1$ becomes 
 \dis{E_s(z^a, \overline{z}^a ; A) \sim |z|^{s} \ll \frac{1}{g_s}\frac{{\cal V}_E}{t_i^\perp}.}
 While the size of some 2-cycle $t_i^\perp$ may be much smaller than ${\cal V}_E^{1/3}$, which indeed is realized in the large volume scenario, it cannot  exceed ${\cal V}_E^{1/3}$ so ${\cal V}_E/t_i^\perp$ is required to be larger than ${\cal V}_E^{2/3}$.
 Therefore, the perturbative control is maintained for any value of $t_i^\perp$ provided  
 \dis{|z|^{s} \ll \frac{{\cal V}_E^{2/3}}{g_s}\sim \frac{1}{g_s}\frac{M_{\rm Pl}}{M_{\rm KK}}\ll \Big(\frac{M_{\rm Pl}}{M_{\rm KK}}\Big)^2,}
where \eqref{eq:KKscale} and \eqref{eq:Inqtau} are  used for the last two expressions.
In order to express this in the distance conjecture-like form, suppose the kinetic term for $z$ is given by $\alpha^2 M_{\rm Pl}^2 |\partial_\mu   z|^2/({\rm Im} z)^{2}$.
  Here $\alpha$ is a dimensionless constant, which is determined by the structure of the tree level K\"ahler potential. 
More precisely, this can be realized when all the moduli (including Re$z$) except for Im$z$ are stabilized to small values, while Im$z$ has a large value :
From  \eqref{eq:KahlerCS},  the kinetic term for Im$z$ is given by
\dis{&M_{\rm Pl}^2\partial_z \partial_{\overline z}K_{\rm cs}  |\partial_\mu   z|^2
=M_{\rm Pl}^2\Big[\frac{\partial_z X \partial_{\overline z}X}{X^2}-\frac{\partial_z \partial_{\overline z} X}{X}\Big] |\partial_\mu   z|^2,
\\ &{\rm where}~~
X=2(F-\overline{F})-(z^a-\overline{z}^a)(\partial_a F+\overline{\partial_a F}),}
So if $F=z^n+\cdots$ and Re$z \simeq 0$, the kinetic term is written as $\alpha^2 M_{\rm Pl}^2 |\partial_\mu   z|^2/({\rm Im} z)^{2}$ where $\alpha^2=n(n-2)/4$ with $n$ being an odd integer larger than $3$.
When $n=2$, $\alpha^2=4/\pi^2$ but for other even $n$, $\alpha^2$ is divergent.
Presumably, the latter case may be allowed when $z$ couples to other moduli so $F$ depends nontrivially on $z$. 
If other terms containing $z$ can be negligible, the geodesic distance of Im$z$ in the field space  is estimated as $\varphi=\alpha \log(|{\rm Im}z|)$.
Then the above inequality can be rewritten as
 \dis{&\varphi \ll 2\frac{\alpha}{s}\log  \Big(\frac{M_{\rm Pl}}{M_{\rm KK}}\Big),
 \quad\quad
  {\rm or}\quad\quad  M_{\rm KK}\ll M_{\rm Pl}e^{-\frac{s}{2\alpha}\varphi}.\label{eq:zMKKbound}}
  In terms of $\Lambda_{\rm sp}^{\rm (KK)}$ and $N_{\rm sp}^{\rm (KK)}$, it reads
 \dis{\Lambda^{\rm (KK)}_{\rm sp}\ll M_{\rm Pl}e^{-(3/8)(s/\alpha)\varphi},\quad\quad N^{\rm (KK)}_{\rm sp} \gg e^{(3/4)(s/\alpha)\varphi}.}

On the other hand, we can infer from \eqref{eq:KKscale} that our condition can be written in terms of $M_s$ instead of $M_{\rm KK}$ :
\dis{|z|^{s} \ll \frac{{\cal V}_E^{2/3}}{g_s} \sim \frac{1}{g_s^{2/3}}\Big(\frac{M_{\rm Pl}}{M_s}\Big)^{4/3} \ll \Big(\frac{M_{\rm Pl}}{M_s}\Big)^2,}
 thus
  \dis{&\varphi \ll 2\frac{\alpha}{ s}\log  \Big(\frac{M_{\rm Pl}}{M_s}\Big),
 \quad\quad
  {\rm or}\quad\quad  M_s\ll M_{\rm Pl}e^{-\frac{s}{2\alpha}\varphi}.\label{eq:zMsbound}} 
    In terms of $\Lambda_{\rm sp}^{(s)}$ and $N_{\rm sp}^{(s)}$, it reads
  \dis{\Lambda_{\rm sp}^{(s)} =M_s \ll M_{\rm Pl}e^{-\frac{s}{2\alpha}\varphi},\quad\quad N_{\rm sp}^{(s)} \gg e^{ (s/\alpha)\varphi}.}

 Meanwhile, from $\delta K_{\rm KK}\sim (g_s/{\cal V}_E^{2/3})\times E_s(z^a, \overline{z}^a; A)$ and $\delta K_{\rm W}\sim (1/{\cal V}_E^{4/3})\times E_s(z^a, \overline{z}^a; A)$, or equivalently, $\delta K_{\rm W}/\delta K_{\rm KK}\sim 1/(g_s {\cal V}_E^{2/3})$, one finds that $\delta K_{\rm W}$ cannot exceed $\delta K_{\rm KK}$ provided $g_s {\cal V}_E^{2/3}>1$, which is nothing more than \eqref{eq:dilcond}, i.e., the condition for the perturbative control by   large  ${\cal V}$.
 This is just a restatement of the fact that in the string frame $\delta K_{\rm KK}$ and $\delta K_{\rm W}$ are of  ${\cal O}(g_s^2{\alpha'}^2)$ and ${\cal O}(g_s^2{\alpha'}^4)$, respectively.
In any case, so far as the perturbative control is maintained, $\delta K_{\rm KK}>\delta K_{\rm W}$ is automatically satisfied.

  We also note that in the large volume scenario \cite{Balasubramanian:2005zx},  the corrected K\"ahler potential given by \eqref{eq:Kahler1} is used to stabilize the K\"ahler moduli.
   For this model to work, the corrections \eqref{eq:CSMCon} must be suppressed compared to the ${\cal O}({\alpha'}^3)$ correction, i.e., $\delta K_{\rm KK} \ll \xi/(g_s^{3/2}{\cal V}_E)$.
 This condition reads
 \dis{E_s(z^a, \overline{z}^a ; A) \sim |z|^{s} \ll \frac{1}{g_s^{5/2}{\cal V}_E^{1/3}}.}
 Replacing ${\cal V}_E$ by the relations given by  \eqref{eq:KKscale}, it can be rewritten as
\dis{\varphi \ll \frac52 \frac{\alpha}{s}|\Phi|-\frac{\alpha}{2s}\log\Big(\frac{M_{\rm Pl}}{M_{\rm KK}}\Big),\quad\quad{\rm or}\quad\quad 
\varphi \ll \frac83 \frac{ \alpha}{ s}|\Phi|-\frac{2\alpha}{3s}\log\Big(\frac{M_{\rm Pl}}{M_s}\Big).}
Since  the value of $\Phi$ is restricted by \eqref{eq:gsMKK} and \eqref{eq:gsMS} for $\delta K_{\rm KK}$ to be suppressed compared to the tree-level K\"ahler potential $-2\log{\cal V}$, these inequalities become
\dis{\varphi \ll 2\frac{\alpha}{s}\log\Big(\frac{M_{\rm Pl}}{M_{\rm KK}}\Big),\quad\quad{\rm or}\quad\quad 
\varphi \ll 2\frac{ \alpha}{ s} \log\Big(\frac{M_{\rm Pl}}{M_s}\Big).}
These bounds coincide with \eqref{eq:zMKKbound} and \eqref{eq:zMsbound}.
In this way, we can obtain various distance conjecture-like bounds on $\varphi$, hence $|z|$, from the condition of the perturbative control.

 \section{Conclusions}
\label{sec:conclusion}

In this article, we investigate the condition that  the perturbative control of string theory model in the framework of the effective supergravity is not lost in both the large volume and the weak coupling expansions, focusing on the field values of the complex structure moduli which has the similar symmetry structure  to the axio-dilaton.
For this purpose, we point out that when the  complex structure moduli (axio-dilaton) couple (couples) to some lattice structure,  the  Sp$(2(h^{2,1}+1))$ (SL$(2, \mathbb{Z})$) symmetry respected by the tree-level K\"ahler potential allows the invariant function which diverges in the large field limit.
If the corrected K\"ahler potential also respects this symmetry, it may contain the function we considered, resulting in the breakdown of the perturbative control in the strict asymptotic regime.
Therefore, the field values of  the complex structure moduli are required to be bounded from above, and these conditions can be written in terms of ${\cal V}$ and $g_s$, or equivalently, the tower mass scale like the KK or the string mass scales.
Then the bounds we obtained can be interpreted as the conditions that  a tower of states decouples from the low energy EFT, just like the distance conjecture.
We note that the similar bound on the dilaton can be obtained, which is nothing more than the restatement of the large string frame volume condition in the language of the Einstein frame.
 
 The existence of the invariant function which  diverges in the large field limit is well supported by the explicitly calculable models.
 Whereas the function is modified by the symmetry breaking effects from, for example, the gauge fields of the brane, this does not affect the large field behavior of the  function.
 Meanwhile, in the explicitly calculable models, the lattice which couples to the moduli is given by that of the KK modes.
 Since the correction to the K\"ahler potential can be obtained from the corrections to the kinetic terms for the moduli, our consideration corresponds to an example of the discussions concerning the effects of a tower of states on the EFT through the correction to the kinetic term, ranging from the emergence proposal \cite{Harlow:2015lma, Heidenreich:2017sim, Grimm:2018ohb, Heidenreich:2018kpg, Castellano:2022bvr} to the constraints on the scale of the non-renormalizable interaction  \cite{Reece:2025thc, Seo:2024zzs, Seo:2025zrr}.
 We also note that the lattice may originate from the quantized 3-form flux, which couples to the complex structure moduli through the Gukov-Vafa-Witten superpotential \cite{Gukov:1999ya}, even though the concrete mechanism is not clear at present.


%


\appendix

  \section{Review on the explicitly calculable examples}
  \label{App:example}

 In this appendix, we review examples  considered in \cite{Berg:2005ja} where  the functional dependence  of $\delta K^{\rm KK}$ on the complex structure modulus $U$ (for the $\mathbb{T}^2$ factor of the internal manifold)  is explicitly calculable.
 From this, we can see that the coupling of $U$ to the  KK modes generates the $U$-dependent part of $\delta K^{\rm KK}$  given by  $E_2(U, \overline{U}; A)$.
 In addition, the explicit dependence of $\delta K^{\rm KK}$ on the K\"ahler moduli can be found as well.
 For our purposes it is sufficient to focus on the simplest example, ${\cal N}=2$ orientifold compactification on $\mathbb{T}^2\times$K3 containing D$9$- and D$5$-branes. 
 Modifications for the model containing  D$7$- and D$3$-branes will be addressed later.
 
 In our example, the complex structure modulus for $\mathbb{T}^2$ is given by
 \dis{U=\frac{1}{G_{44}}(G_{45}+i\sqrt{G}),}
such that the metric for $\mathbb{T}^2$ in the string frame is written as
 \dis{G_{mn}=\frac{\sqrt{G}}{U_2} 
 \left(
\begin{array}{cc}
1 & U_1 \\
U_1 & |U|^2
\end{array}\right),
\label{eq:T2metric} }
 where the indices $m, n$ run over $4,5$.
 Here   Re$U$ and Im$U$ are denoted by $U_1$ and $U_2$, respectively, and other moduli as well as the complex parameters will be denoted in the same way.
 The K\"ahler moduli for the overall volume and $\mathbb{T}^2$ are given by
 \dis{&S=\frac{1}{\sqrt{8\pi^2}}(C+ie^{-\Phi}\sqrt{G}{\cal V}_{\rm K3}),
\quad\quad S'=\frac{1}{\sqrt{8\pi^2}}(C_{45}+ie^{-\Phi}\sqrt{G}),\label{eq:Kahlerex}}
respectively, where $C$ is defined from the RR 2-form $C_2$ by $dC=*_4 dC_2$ and $C_{45}$ is the component of $C_2$ along $\mathbb{T}_2$.
Moreover, there are 10-dimensional Abelian vectors of the D-brane stack $i$, the components along $\mathbb{T}^2$ of which are denoted by $a_m^i$.
The 4-dimensional effective action is given by the   dimensional reduction of the sum of the type-I supergravity action, the DBI action for the D$9$-brane, and the DBI action for the D$5$-brane : 
\dis{S= \frac{\ell_s^6}{\kappa_{10}^2}\int d^4 \sqrt{-g_4}\big[{\cal L}_{\rm SUGRA}+{\cal L}_{{\rm D}9}+{\cal L}_{{\rm D}5}\big],}
where
\dis{&{\cal L}_{\rm SUGRA}=\frac12{\cal R}+\frac{\partial_\mu S\partial^\mu \overline{S}}{(S-\overline{S})^2}+\frac{\big|\partial_\mu S'+\frac{1}{8\pi}\sum_i N_i(a_4^i\partial_\mu a_5^i-a_5^i\partial_\mu a_4^i)\big|^2}{(S'-\overline{S'})^2}+\frac{\partial_\mu U\partial^\mu \overline{U}}{(U-\overline{U})^2}+\cdots,
\\
&{\cal L}_{{\rm D}9}=\frac{1}{4\pi}\frac{\sum_i N_i\big|U \partial_\mu a_4^i-\partial_\mu a_5^i\big|^2}{(U-\overline{U})(S'-\overline{S'})}-\frac14\frac{\kappa_{10}^2}{\ell_s^6}S_2 {\rm tr}F_{{\rm D}9}^2+\cdots,
\\
&{\cal L}_{{\rm D}5}=-\frac14\frac{\kappa_{10}^2}{\ell_s^6}S'_2 {\rm tr}F_{{\rm D}5}^2+\cdots. \label{eq:L}}
While the SL$(2)$ symmetry is respected by the kinetic term for $U$, as can be seen in the first term of ${\cal L}_{{\rm D}9}$, it is broken by the kinetic term for $a^i_m$ : under $U \to (a U+ b)/(cU+d)$, the combination $|U \partial_\mu a_4^i-\partial_\mu a_5^i|^2/(U-\overline{U})$ transforms to $|U \partial_\mu {\tilde a}_4^i-\partial_\mu {\tilde a}_5^i|^2/(U-\overline{U})$ where ${\tilde a}^i_4= a a_4^i -c a_5^i$ and ${\tilde a}^i_5= -b a_4^i +d a_5^i$.
The kinetic terms in the action, together with their tree-level corrections can be obtained from the tree-level K\"ahler potential
\dis{K_0=-\log\big[(S-\overline{S})(S'-\overline{S'})(U-\overline{U})\big].}
 
 The 1-loop correction to the K\"ahler potential can be read off from the 1-loop corrections to the kinetic terms for the moduli.
 Here we consider how the kinetic term for Im$S'=S'_2$ is corrected.
 Whereas we need to calculate the annulus, the M\"obius strip, and the Klein bottle contributions, since they are obtained from `covering tori' through appropriate involutions \cite{Antoniadis:1996vw}, for our purpose it suffices to sketch how $E_2(U, \overline{U}; A)$ is obtained  when the world-sheet topology is given by the torus.
 For technical details, the readers are invited to refer to \cite{Berg:2005ja, Antoniadis:1996vw}.
 On the torus with modulus $\tau=\tau_1+i\tau_2$, 
 \footnote{Whereas we have used $\tau$ to denote the axio-dilaton, $\tau$ in this appendix before \eqref{eq:deltaK} represents the modulus for the torus, the world-sheet topology of the string loop correction, following the convention. 
 This torus must be distinguished from the internal $\mathbb{T}^2$ geometry along the $4, 5$ directions. }
 the target space vectors along the internal directions $m=4, 5$ are expanded as $X^m=x_0^m+\sqrt{\alpha'} n^m (w_2/\tau_2)+\cdots$, where $n^m \in \mathbb{Z}$ and the string world-sheet coordinates are given by $w_i \in [0, 2\pi)$ ($i=1,2$ ; $z=e^{-i(w_1+i w_2)}$).
 The momentum, or equivalently, the KK mode,  is quantized such that as $w_2 \to w_2+2\pi \tau_2$, $X^m$ transforms to $X^m+2\pi n^m\sqrt{\alpha'}$.
 Then the world-sheet action includes
 \dis{S_{\rm P} \ni-\frac{1}{4\pi\alpha'}\int_\Sigma d^2w\sqrt{h} G_{mn}\dot{X}^m\dot{X}^n =-\frac{\pi}{\tau_2} G_{pq}n^p n^q= -\frac{\pi}{\tau_2}\frac{\sqrt{G}}{U_2}|n^4 +\overline{U}n^5|^2,}
 where $G_{mn}$ is given by \eqref{eq:T2metric} and $\int_\Sigma d^2w \sqrt{h}=4\pi^2 \tau_2$.
 It is often convenient to introduce the complex coordinate
 \dis{Z=\sqrt{\frac{\sqrt{G}}{2U_2}}(X^4+\overline{U}X^5),\label{eq:Z}}
  in terms of which the action can be rewritten in the  quite simple form : 
 \dis{f_1(\sqrt{G}, U)\equiv G_{mn}\partial X^m \overline{\partial}X^n=\partial \overline{Z} \overline{\partial}Z+\partial Z \overline{\partial}\overline{Z}.}
 For the annulus and the M\"obius strip, the world-sheet topology contains the boundary, so  we need to add the boundary action in the form of
 \dis{S_B=\frac{i}{\sqrt{\alpha'}}\oint_{\partial\Sigma}a_m^i \dot{X}^m \ni 2\pi i n^m a^i_m = 2\pi i (n^4 a_4^i+n^5 a_5^i).\label{eq:SB}}
 
 In order to find the loop corrections to the kinetic term for $S'_2$, we need to compute the 2-point correlation function $\langle V_{S'_2}V_{S'_2}\rangle$.
 Here   $V_{S'_2}$ is the vertex operator for $S'_2$ given by $V_{S'_2}\sim \frac{i}{S'-\overline{S'}}V_{Z\overline{Z}}$, which can be read off from
 \footnote{To obtain this relation, we use $\partial_{\sqrt{G}}f_1=f_1/\sqrt{G}$, $\partial_U f_1=(i/U_2) f_1$ and also $\partial_{S'_2}\sqrt{G}=\sqrt2 \pi e^\Phi$.
 For the last relation  we first take derivative of the expression
\dis{\sqrt{G}=-i\sqrt2\pi\Big[\frac{(S-\overline{S})(S'-\overline{S'})}{e^{-2\Phi}{\cal V}_{\rm K3}}\Big]^{1/2}}
with respect to $S'_2$ then simplify  it using $\sqrt{2i}e^\Phi=[(S-\overline{S})/(e^{-2\Phi}{\cal V}_{\rm K3} S'_2)]^{1/2}$ \cite{Berg:2005ja}.
Applying $e^\Phi/\sqrt{G}=[\sqrt2 \pi (S'-\overline{S'})]^{-1}$ in addition, we reach \eqref{eq:dSdS2}.}
  \dis{\frac{\delta S_P}{\delta S'_2}=\partial_{S'_2}\sqrt{G} \partial_{\sqrt{G}}S_P=\frac{1}{2\pi\alpha'}\int_{\Sigma} d^2 z \frac{i}{S'-\overline{S'}}(\partial \overline{Z} \overline{\partial}Z+\partial Z \overline{\partial}\overline{Z}). \label{eq:dSdS2}}
Hence, $\langle V_{S'_2}V_{S'_2}\rangle$ is written as 
\dis{\langle  V_{S'_2} V_{S'_2}\rangle \sim \frac{1}{(S'-\overline{S'})^2 }\langle V_{Z\overline{Z}} V_{Z\overline{Z}}\rangle.\label{eq:VsVs}}
When $V_{Z\overline{Z}}$ is of the 0-picture, it contains $\overline{\partial}Z$, $\overline{\partial}\overline{Z}$, etc, then $\langle V_{Z\overline{Z}} V_{Z\overline{Z}}\rangle$ is given by  a term $\langle \overline{\partial}Z (w) \overline{\partial}\overline{Z}(w') \rangle$ multiplied by the  contributions from the fermions we are not interested in.
In this case, the function   $E_2(U, \overline{U}; A)$ comes out from the following two contributions.

First, from \eqref{eq:Z} one immediately finds that $\langle \overline{\partial}Z \overline{\partial}\overline{Z} \rangle$ contains a term proportional to $\frac{\sqrt{G}}{2U_2 \tau_2^2}$ $|n^4+\overline{U}n^5|^2$.
Then $\langle V_{Z\overline{Z}} V_{Z\overline{Z}}\rangle$ is proportional to
\dis{ \sqrt{G}\int\frac{d\tau_2}{\tau_2^4}\int d^2w \sqrt{h} \int d^2w' \sqrt{h} \sum_{(n^4, n^5) \in \mathbb{Z}}{\rm tr}\Big[e^{-\frac{\pi}{\tau_2}G_{pq}n^pn^q+2\pi i n^p a^i_p}\frac{\sqrt{G}}{2U_2 \tau_2^2} |n^4+\overline{U}n^5|^2\Big],}
where $\sqrt{h}=\tau_2$.
We note that whereas $d\tau_2/\tau_2^2$ is a modular invariant integral measure, it is additionally multiplied by $1/\tau_2^2$ coming from   the contribution of the 4-dimensional momentum, $\int d^4 x e^{-2\pi\tau_2\alpha' p^2}$.
Integral over $\tau_2$ can be easily calculated by the change of the variable  $\tau_2 =1/l$, giving
\dis{\frac{1}{\sqrt{G}}E_2(U, \overline{U} ; A)=\frac{1}{\sqrt{G}}\sum_{(n^4, n^5)\ne (0,0)}e^{2\pi i n^p a^i_p}\frac{U_2^2}{|n^4+Un^5|^4}.\label{eq:loopE2}}
where we omitted irrelevant factors and   $A$  indicates the combination $A^i=Ua_4^i-a_5^i$.
When $a_m^i=0$, $E_2(U, \overline{U}; A)$ becomes the SL$(2)$ invariant function $E_2(U, \overline{U})$ given by \eqref{eq:E(z)}.
Meanwhile, for $a_m^i\ne 0$, $E_2(U, \overline{U}; A)$ is not invariant under the SL$(2)$ transformation $U\to (aU+b)/(cU+d)$ as the exponent in the summand becomes $2\pi i m^p\tilde{a}_p^i$ where ${\tilde a}^i_4= a a_4^i -c a_5^i$ and ${\tilde a}^i_5= -b a_4^i +d a_5^i$.
This indeed is what expected from the structure of the kinetic term for $a_m^i$.

Second, we observe that the Green's function on the torus contains a term (see, for example, Sec. 7.2 of \cite{Polchinski:1998rq})
\dis{\langle Z(w)\overline{Z}(w')\rangle \sim-\frac{\alpha'}{2}\log\Big|\frac{2\pi}{\vartheta'_1(0|\tau)}\vartheta_1\Big(\frac{w_1-w_2}{2\pi}\Big|\tau\Big)\Big|^2 +\alpha'\frac{(w_2-w'_2)^2}{4\pi\tau_2}.}
Of particular interest is the second term, which neutralizes the background charge distribution in the compact space.
Its contribution to $\langle V_{Z\overline{Z}} V_{Z\overline{Z}}\rangle$ is proportional to
\dis{\sqrt{G} \int\frac{d\tau_2}{\tau_2^4}\int d^2w \sqrt{h} \int d^2w' \sqrt{h} \sum_{(n^4, n^5) \in \mathbb{Z}}{\rm tr}\Big[e^{-\frac{\pi}{\tau_2}G_{pq}n^pn^q+2\pi i n^p a^i_p}\frac{\alpha'}{4\pi \tau_2}\Big],}
and the integration over $\tau_2$ in the same way as the first case gives $E_2(U, \overline{U} ; A)/\sqrt{G}$, upto irrelevant factors.

Now the factor $1/\sqrt{G}$ in \eqref{eq:loopE2} can be rewritten as  $e^{-\Phi}[\sqrt2 \pi (S'-\overline{S'})]^{-1}$, where $e^{-\Phi}$ can be substituted by $-i (e^{-2\Phi}\sqrt{G}{\cal V}_{\rm K3}/[\sqrt2\pi (S-\overline{S})])$ (see \eqref{eq:Kahlerex}).
Here the combination $e^{-2\Phi}\sqrt{G}{\cal V}_{\rm K3}$ corresponds to the part of the prefactor $\sqrt{-G_{10}}e^{-2\Phi}$ in the effective supergravity action.
Therefore, combined with $(S'-\overline{S'})^{-2}$ in \eqref {eq:VsVs}, the correction to the kinetic term for $S_2'$, or equivalently, $K_{S'\overline{S'}}$ is written as
\dis{K_{S'\overline{S'}} \sim \frac{{\cal E}_2(U, \overline{U} ; A)}{(S'-\overline{S'})^3 (S-\overline{S})},}
where ${\cal E}_2(U, \overline{U}; A)$ is the linear combination of  $E_2(U, \overline{U} ; A)$s.
More explicitly,  it is given by
\dis{{\cal E}_2(U, \overline{U}; A)=-4\sum_{i,j}N_iN_i\big[&E_2(U, \overline{U}; A_i-A_j)+E_2(U, \overline{U}; -A_i+A_j)
\\
&-E_2(U, \overline{U}; A_i+A_j)-E_2(U, \overline{U}; -A_i-A_j)\big]
\\
+64\sum_i N_i[&E_2(U, \overline{U}; A_i)+E_2(U, \overline{U}; -A_i)]
\\
-4\sum_iN_i[&E_2(U, \overline{U}; 2A_i)+E_2(U, \overline{U}; -2A_i)].}
As can be seen from \eqref{eq:E2zz}, in the limit $|U|\to\infty$, the leading term of $E_2(U, \overline{U}; A)$ is independent of $A$ but given by $(\pi^4/45)U_2^2$, just like $E_2(U, \overline{U})$.
This leads to ${\cal E}_2(U, \overline{U}; A) \simeq (\sum_i N)\times (8\pi^4/3)U_2^2$ in the same limit. 
Finally, from the correction to $K_{S'\overline{S'}}$, one finds that  the correction to the K\"ahler potential is 
\dis{\delta K=\sum_i \frac{c {\cal E}_2(U, \overline{U}; A)}{(S-\overline{S})(S'-\overline{S'})},}
where the coefficient $c$ turns out to be $1/(128\pi^6)$.

On the other hand, in the ${\cal N}=1$ orientifold compactification on $\mathbb{T}^6/(\mathbb{Z}_2\times \mathbb{Z}_2)$ containing D$9$- and D$5$-branes, the T-duality allows additional contributions, which will be interpreted as $\delta K_W$ in \eqref{eq:deltaKs}. 
Then the total 1-loop correction to the K\"ahler potential is given by
\dis{\delta K=\frac{c}{2}\sum_{I=1}^3 \frac{{\cal E}^{\mathbb{Z}_2^2}_2(U, \overline{U}; A)}{(S-\overline{S})(T^I-\overline{T}^I)}+\frac{c}{2}\sum_{I=1}^3 \frac{{\cal E}^{\mathbb{Z}_2^2}_2(U, \overline{U^I}; 0)}{(T^J-\overline{T}^J)(T^K-\overline{T}^K)}\Big|_{I\ne J\ne K},}
where $I, J, K$ indicate the three tori comprising $\mathbb{T}^6$, the K\"ahler moduli for which are given by  $T^I$.
The function ${\cal E}^{\mathbb{Z}_2^2}(U, \overline{U}; A)$ is given by
 \dis{{\cal E}_2^{\mathbb{Z}_2^2}(U, \overline{U}; A)=&128\sum_iN_i[E_2(U, \overline{U};A)+E_2(U, \overline{U};-A)]
 \\
 &-8\sum_iN_i[E_2(U, \overline{U};2A)+E_2(U, \overline{U};-2A)].}
Now, when we consider D$7$- and D$3$-branes rather than D$9$- and D$5$-branes, the corrections can be obtained from those in the case of D$9$- and D$5$-branes  by taking T-dualities along all internal cycles.
Then $S$ becomes the axio-dilaton $\tau$, and $T^I_2$ becomes the K\"ahler modulus for the 4-cycle $\rho_2^I \sim \partial{\cal V}/\partial T_2^I$, thus
 \dis{\delta K=\frac{c}{2}\sum_{I=1}^3 \frac{{\cal E}_2^{\mathbb{Z}_2^2}(U^I, \overline{U^I}; A)}{(\tau-\overline{\tau})(\rho^I-\overline{\rho}^I)}+\frac{c}{2}\sum_{I=1}^3 \frac{{\cal E}_2^{\mathbb{Z}_2^2}(U^I, \overline{U^I}; 0)}{(\rho^J-\overline{\rho}^J)(\rho^K-\overline{\rho}^K)}\Big|_{I\ne J\ne K}.\label{eq:deltaK}}
In the same way as \eqref{eq:Kahlerex}, in the string frame, Im$\rho^I$ is given by $e^{-\Phi}$ times the volume of the 4-cycle in units of the string length.
Therefore, two terms in $\delta K$ are of ${\cal O}(g_s^2\alpha^2)$ and ${\cal O}(g_s^2\alpha^4)$, respectively.
On the other hand, in the Einstein frame, the factor $e^{-\Phi}=e^{-4\times\frac14\Phi}$ in Im$\rho^I$ is absent, thus two terms are regarded as of ${\cal O}(g_s^2{\cal V}_E^{-2/3})$ and ${\cal O}({\cal V}_E^{-4/3})$, respectively.



\begin{thebibliography}{99}

\small


\bibitem{Ibanez:2012zz}
L.~E.~Ibanez and A.~M.~Uranga,
``String theory and particle physics: An introduction to string phenomenology,''
Cambridge University Press, 2012.

\bibitem{Tomasiello:2022dwe}
A.~Tomasiello,
``Geometry of String Theory Compactifications,''
Cambridge University Press, 2022.

\bibitem{Conlon:2005ki}
J.~P.~Conlon, F.~Quevedo and K.~Suruliz,
JHEP \textbf{08} (2005), 007
[arXiv:hep-th/0505076 [hep-th]].

\bibitem{Ooguri:2006in}
H.~Ooguri and C.~Vafa,
Nucl. Phys. B \textbf{766} (2007), 21-33
[arXiv:hep-th/0605264 [hep-th]].

\bibitem{Lee:2019xtm}
S.~J.~Lee, W.~Lerche and T.~Weigand,
JHEP \textbf{02} (2022), 096
[arXiv:1904.06344 [hep-th]].

\bibitem{Lee:2019wij}
S.~J.~Lee, W.~Lerche and T.~Weigand,
JHEP \textbf{02} (2022), 190
[arXiv:1910.01135 [hep-th]].

\bibitem{Grimm:2018ohb}
T.~W.~Grimm, E.~Palti and I.~Valenzuela,
JHEP \textbf{08} (2018), 143
[arXiv:1802.08264 [hep-th]].

\bibitem{Grimm:2018cpv}
T.~W.~Grimm, C.~Li and E.~Palti,
JHEP \textbf{03} (2019), 016
[arXiv:1811.02571 [hep-th]].

\bibitem{Schmid:1973cdo}
W.~Schmid,
Invent. Math. \textbf{22} (1973) no.3, 211-319

\bibitem{Carrani:1986}
E.~Cattani, A.~Kaplan and W.~Schmid,
 Ann. of Math. \textbf{123} (1986) 457.


\bibitem{Grimm:2019ixq}
T.~W.~Grimm, C.~Li and I.~Valenzuela,
JHEP \textbf{06} (2020), 009
[erratum: JHEP \textbf{01} (2021), 007]
[arXiv:1910.09549 [hep-th]].

\bibitem{Bena:2020xrh}
I.~Bena, J.~Bl\r{a}b\"ack, M.~Gra\~na and S.~L\"ust,
JHEP \textbf{11} (2021), 223
[arXiv:2010.10519 [hep-th]].

\bibitem{Bena:2021wyr}
I.~Bena, J.~Bl\r{a}b\"ack, M.~Gra\~na and S.~L\"ust,
Adv. Appl. Clifford Algebras \textbf{32} (2022) no.1, 7
[arXiv:2103.03250 [hep-th]].



\bibitem{Dasgupta:1999ss}
K.~Dasgupta, G.~Rajesh and S.~Sethi,
JHEP \textbf{08} (1999), 023
[arXiv:hep-th/9908088 [hep-th]].

\bibitem{Giddings:2001yu}
S.~B.~Giddings, S.~Kachru and J.~Polchinski,
Phys. Rev. D \textbf{66} (2002), 106006
[arXiv:hep-th/0105097 [hep-th]].

\bibitem{Grana:2005jc}
M.~Grana,
Phys. Rept. \textbf{423} (2006), 91-158
[arXiv:hep-th/0509003 [hep-th]].


\bibitem{Douglas:2006es}
M.~R.~Douglas and S.~Kachru,
Rev. Mod. Phys. \textbf{79} (2007), 733-796
[arXiv:hep-th/0610102 [hep-th]].

\bibitem{Blumenhagen:2006ci}
R.~Blumenhagen, B.~Kors, D.~Lust and S.~Stieberger,
Phys. Rept. \textbf{445} (2007), 1-193
[arXiv:hep-th/0610327 [hep-th]].

\bibitem{Plauschinn:2021hkp}
E.~Plauschinn,
JHEP \textbf{02} (2022), 206
[arXiv:2109.00029 [hep-th]].

\bibitem{Lust:2021xds}
S.~L\"ust,
 arXiv preprint (2021)  [arXiv:2109.05033 [hep-th]].

\bibitem{Grimm:2021ckh}
T.~W.~Grimm, E.~Plauschinn and D.~van de Heisteeg,
JHEP \textbf{03} (2022), 117
[arXiv:2110.05511 [hep-th]].

\bibitem{Grana:2022dfw}
M.~Gra\~na, T.~W.~Grimm, D.~van de Heisteeg, A.~Herraez and E.~Plauschinn,
JHEP \textbf{08} (2022), 237
[arXiv:2204.05331 [hep-th]].

\bibitem{Tsagkaris:2022apo}
K.~Tsagkaris and E.~Plauschinn,
JHEP \textbf{03} (2023), 049
[arXiv:2207.13721 [hep-th]].


\bibitem{Vafa:1996xn}
C.~Vafa,
Nucl. Phys. B \textbf{469} (1996), 403-418
[arXiv:hep-th/9602022 [hep-th]].

\bibitem{Schwarz:1995dk}
J.~H.~Schwarz,
Phys. Lett. B \textbf{360} (1995), 13-18
[erratum: Phys. Lett. B \textbf{364} (1995), 252]
[arXiv:hep-th/9508143 [hep-th]].

\bibitem{Aspinwall:1995fw}
P.~S.~Aspinwall,
Nucl. Phys. B Proc. Suppl. \textbf{46} (1996), 30-38
[arXiv:hep-th/9508154 [hep-th]].

\bibitem{ValeixoBento:2025yhz}
B.~Valeixo Bento and M.~Montero,
  arXiv preprint (2025)  [arXiv:2507.02037 [hep-th]].

\bibitem{Chen:2025rkb}
S.~Chen, D.~van de Heisteeg and C.~Vafa,
JHEP \textbf{07}, 258 (2025)
[arXiv:2503.16599 [hep-th]].

\bibitem{Antoniadis:1997eg}
I.~Antoniadis, S.~Ferrara, R.~Minasian and K.~S.~Narain,
Nucl. Phys. B \textbf{507} (1997), 571-588
[arXiv:hep-th/9707013 [hep-th]].

\bibitem{Lust:2022mhk}
S.~L\"ust and M.~Wiesner,
JHEP \textbf{12} (2023), 029
[arXiv:2211.05128 [hep-th]].

\bibitem{Coudarchet:2023mmm}
T.~Coudarchet, F.~Marchesano, D.~Prieto and M.~A.~Urkiola,
JHEP \textbf{08} (2023), 016

[arXiv:2304.04789 [hep-th]].


\bibitem{Lanza:2024uis}
S.~Lanza and A.~Westphal,
JHEP \textbf{05}, 071 (2025)
[arXiv:2412.12253 [hep-th]].

\bibitem{Balasubramanian:2005zx}
V.~Balasubramanian, P.~Berglund, J.~P.~Conlon and F.~Quevedo,
JHEP \textbf{03} (2005), 007
[arXiv:hep-th/0502058 [hep-th]].

\bibitem{Junghans:2022exo}
D.~Junghans,
Nucl. Phys. B \textbf{990} (2023), 116179
[arXiv:2201.03572 [hep-th]].

\bibitem{ValeixoBento:2023nbv}
B.~Valeixo Bento, D.~Chakraborty, S.~Parameswaran and I.~Zavala,
JHEP \textbf{11} (2023), 075
[arXiv:2306.07332 [hep-th]].

\bibitem{Gao:2022fdi}
X.~Gao, A.~Hebecker, S.~Schreyer and V.~Venken,
JHEP \textbf{07} (2022), 056
[arXiv:2202.04087 [hep-th]].



\bibitem{Cicoli:2024bxw}
M.~Cicoli, A.~Grassi, O.~Lacombe and F.~G.~Pedro,
JHEP \textbf{06}, 090 (2025)
[arXiv:2412.08723 [hep-th]].



\bibitem{Berg:2005ja}
M.~Berg, M.~Haack and B.~Kors,
JHEP \textbf{11} (2005), 030
[arXiv:hep-th/0508043 [hep-th]].


\bibitem{Berg:2005yu}
M.~Berg, M.~Haack and B.~Kors,
Phys. Rev. Lett. \textbf{96} (2006), 021601
[arXiv:hep-th/0508171 [hep-th]].

\bibitem{Becker:2002nn}
K.~Becker, M.~Becker, M.~Haack and J.~Louis,
JHEP \textbf{06} (2002), 060
[arXiv:hep-th/0204254 [hep-th]].

\bibitem{Green:1997tv}
M.~B.~Green and M.~Gutperle,
Nucl. Phys. B \textbf{498} (1997), 195-227
[arXiv:hep-th/9701093 [hep-th]].
 


\bibitem{Gross:1986iv}
D.~J.~Gross and E.~Witten,
Nucl. Phys. B \textbf{277} (1986), 1

\bibitem{Arkani-Hamed:2006emk}
N.~Arkani-Hamed, L.~Motl, A.~Nicolis and C.~Vafa,
JHEP \textbf{06} (2007), 060
[arXiv:hep-th/0601001 [hep-th]].

\bibitem{Harlow:2022ich}
D.~Harlow, B.~Heidenreich, M.~Reece and T.~Rudelius,
Rev. Mod. Phys. \textbf{95} (2023) no.3, 3
[arXiv:2201.08380 [hep-th]].

\bibitem{Veneziano:2001ah}
G.~Veneziano,
JHEP \textbf{06} (2002), 051
[arXiv:hep-th/0110129 [hep-th]].

\bibitem{Dvali:2007hz}
G.~Dvali,
Fortsch. Phys. \textbf{58} (2010), 528-536
[arXiv:0706.2050 [hep-th]].

\bibitem{Dvali:2007wp}
G.~Dvali and M.~Redi,
Phys. Rev. D \textbf{77} (2008), 045027
[arXiv:0710.4344 [hep-th]].

\bibitem{Dvali:2009ks}
G.~Dvali and D.~Lust,
Fortsch. Phys. \textbf{58} (2010), 505-527
[arXiv:0912.3167 [hep-th]].

\bibitem{Dvali:2010vm}
G.~Dvali and C.~Gomez,
  arXiv preprint (2010)  [arXiv:1004.3744 [hep-th]].

\bibitem{Dvali:2012uq}
G.~Dvali, C.~Gomez and D.~Lust,
Fortsch. Phys. \textbf{61} (2013), 768-778
[arXiv:1206.2365 [hep-th]].


\bibitem{Castellano:2021mmx}
A.~Castellano, A.~Herr\'aez and L.~E.~Ib\'a\~nez,
JHEP \textbf{08} (2022), 217
[arXiv:2112.10796 [hep-th]].

\bibitem{Kani:1989im}
I.~Kani and C.~Vafa,
Commun. Math. Phys. \textbf{130} (1990), 529-580

\bibitem{Berg:2007wt}
M.~Berg, M.~Haack and E.~Pajer,
JHEP \textbf{09} (2007), 031
[arXiv:0704.0737 [hep-th]].

\bibitem{Cicoli:2007xp}
M.~Cicoli, J.~P.~Conlon and F.~Quevedo,
JHEP \textbf{01} (2008), 052
[arXiv:0708.1873 [hep-th]].



\bibitem{Harlow:2015lma}
D.~Harlow,
JHEP \textbf{01} (2016), 122
[arXiv:1510.07911 [hep-th]].

\bibitem{Heidenreich:2017sim}
B.~Heidenreich, M.~Reece and T.~Rudelius,
Eur. Phys. J. C \textbf{78} (2018) no.4, 337
[arXiv:1712.01868 [hep-th]].

\bibitem{Heidenreich:2018kpg}
B.~Heidenreich, M.~Reece and T.~Rudelius,
Phys. Rev. Lett. \textbf{121} (2018) no.5, 051601
[arXiv:1802.08698 [hep-th]].

\bibitem{Castellano:2022bvr}
A.~Castellano, A.~Herr\'aez and L.~E.~Ib\'a\~nez,
JHEP \textbf{06} (2023), 047
[arXiv:2212.03908 [hep-th]].


\bibitem{Reece:2025thc}
M.~Reece,
JHEP \textbf{07}, 130 (2025)
[arXiv:2406.08543 [hep-ph]].


\bibitem{Seo:2024zzs}
M.~S.~Seo,
JHEP \textbf{11} (2024), 082
[arXiv:2407.16156 [hep-th]].


\bibitem{Seo:2025zrr}
M.~S.~Seo,
  arXiv preprint (2025)  [arXiv:2501.02682 [hep-th]].

\bibitem{Gukov:1999ya}
S.~Gukov, C.~Vafa and E.~Witten,
Nucl. Phys. B \textbf{584} (2000), 69-108
[erratum: Nucl. Phys. B \textbf{608} (2001), 477-478]
[arXiv:hep-th/9906070 [hep-th]].

\bibitem{Antoniadis:1996vw}
I.~Antoniadis, C.~Bachas, C.~Fabre, H.~Partouche and T.~R.~Taylor,
Nucl. Phys. B \textbf{489} (1997), 160-178
[arXiv:hep-th/9608012 [hep-th]].

\bibitem{Polchinski:1998rq}
J.~Polchinski,
``String theory. Vol. 1: An introduction to the bosonic string,''
Cambridge University Press, 2007.

\end{thebibliography}
\end{document}